\documentclass[article]{jss}
\usepackage[utf8]{inputenc}

\author{
Jin Zhu\\Sun Yat-Sen University \And Wenliang Pan\\Sun Yat-Sen University \And Wei Zheng\\University of Tennessee \And Xueqin Wang\\Sun Yat-Sen University
}
\title{\pkg{Ball}: An \proglang{R} package for detecting distribution difference and association in metric spaces}

\Plainauthor{Jin Zhu, Wenliang Pan, Wei Zheng, Xueqin Wang}

\Plaintitle{Ball: An R package for detecting distribution difference and association in metric spaces}
\Shorttitle{Ball: Statistical Inference in Metric Spaces via Ball Test Statistics}

\Abstract{
The rapid development of modern technology has created many complex datasets in non-linear spaces, while most of the statistical hypothesis tests are only available in Euclidean or Hilbert spaces. 
To properly analyze the data with more complicated structures, efforts have 
been made to solve the fundamental test problems in more general spaces \citep{lyons2013distance, pan2018ball, pan2018bcov}. 
In this paper, we introduce a publicly available \proglang{R} package \pkg{Ball} for the comparison of multiple distributions and the test of mutual independence in metric spaces, which extends the test procedures for the equality of two distributions \citep{pan2018ball} and the independence of two random objects \citep{pan2018bcov}. 
The \pkg{Ball} package is computationally efficient since several novel algorithms as well as engineering techniques are employed in speeding up the Ball test procedures. 
Two real data analyses and diverse numerical studies have been performed, and the results certify that the \pkg{Ball} package can detect various distribution differences and complicated dependences in complex datasets, e.g., directional data and symmetric positive definite matrix data.
}

\Plainkeywords{K-sample test, test of mutual independence, Ball Divergence, Ball Covariance, metric space}

\Address{
    Jin Zhu\\
  Department of Statistical Science, School of Mathematics \\
  Southern China Center for Statistical Science \\
  Sun Yat-Sen University \\
  510275 Guangzhou, GD, China\\
  E-mail: \email{zhuj37@mail.sysu.edu.cn}\\

      Wenliang Pan\\
  Department of Statistical Science, School of Mathematics \\
  Southern China Center for Statistical Science \\
  Sun Yat-Sen University \\ 
  510275 Guangzhou, GD, China\\
  E-mail: \email{panwliang@mail.sysu.edu.cn}\\

      Wei Zheng\\
  Department of Business analytics and Statistics \\
  University of Tennessee \\
  Knoxville, Tennessee 37996, United States of America\\
  E-mail: \email{wzheng9@utk.edu}\\

      Xueqin Wang\\
  Department of Statistical Science, School of Mathematics \\
  Southern China Center for Statistical Science \\
  Sun Yat-Sen University \\
  510275 Guangzhou, GD, China\\
  E-mail: \email{wangxq88@mail.sysu.edu.cn}\\

  }

\usepackage{amsmath} 
\usepackage{multirow} 
\usepackage{booktabs}
\usepackage{graphicx} 
\usepackage{epstopdf}
\usepackage{algorithm} 
\usepackage{algorithmic}

\begin{document}

\newcommand{\bd}{\text{BD}}
\newcommand{\stbd}{\text{BD}_{N}}
\newcommand{\sksbd}{\text{BD}^{\mathcal{S}_{K}}_{N}}
\newcommand{\skmsbd}{\text{BD}^{\mathcal{MS}_{K}}_{N}}
\newcommand{\skmbd}{\text{BD}^{\mathcal{M}_{K}}_{N}}
\newcommand{\bcov}{\text{BCov}_{\omega}}
\newcommand{\sbcov}{\text{BCov}_{\omega, N}^{K}}
\newcommand{\scbcov}{\text{BCov}_{N}^{K}}
\newcommand{\spbcov}{\text{BCov}_{\Delta, N}^{K}}
\newcommand{\schibcov}{\text{BCov}_{{\chi}^2, N}^{K}}
\newcommand{\skbcov}{\text{BCov}_{\omega, N}^{K}}
\newcommand{\skcbcov}{\text{BCov}_{N}^{K}}
\newcommand{\skpbcov}{\text{BCov}_{\Delta, N}^{K}}
\newcommand{\skchibcov}{\text{BCov}_{{\chi}^2, N}^{K}}
\newcommand{\sbcor}{\text{BCor}_{\omega, N}^{K}}
\newcommand{\skbcor}{\text{BCor}_{\omega, N}^{K}}
\newcommand{\bcor}{\text{BCor}}
\newcommand{\Cov}{\text{cov}}
\newcommand{\Var}{\text{var}}
\newcommand{\coef}{\mathbf{c}}

\section{Introduction}\label{introduction}

With the advanced modern instruments such as the Doppler shift acoustic radar, functional magnetic resonance imaging (fMRI) apparatus, and Heidelberg retina tomograph device, a large number of complex datasets are being collected for contemporary scientific research. 
For example, to investigate whether the wind directions of two places are distinct, meteorologists measure the wind directions by colatitude and longitude coordinates on the earth. Another typical example arises in biology. By using fMRI data, biologists are able to study the association between the brain connectivity and the age. 
Although these complex datasets are potentially useful for the progress of scientific research, their various and complicated structures challenge testing the equality of distributions and testing the mutual independence of random objects, two fundamental problems of statistical inference.
The two problems are generally named as the \(K\)-sample test problem and the test of mutual independence problem, which we reconsider in general metric spaces here.

In the literature, a large number of methods have been developed to address these two problems. Correspondingly, there are many functions currently available in \proglang{R} \citep{RCT2017R}, including \code{oneway.test}, \code{kruskal.test} and \code{cor.test} from package \pkg{stats}, \code{ad.test} from package \pkg{kSamples} \citep{fritz2018ksamples}, \code{tauStarTest} from package \pkg{TauStar} \citep{luca2019taustar}, \code{HellCor} from package \pkg{HellCor} \citep{geenens2018hellinger}, \code{hotelling.test} from package \pkg{Hotelling} \citep{Curran2017Hotelling}, \code{coeffRV} from package \pkg{FactoMineR} \citep{le2008factominer}, \code{kmmd} from package \pkg{kernlab} \citep{Karatzoglou2004}, \code{hsic.test} from package \pkg{kpcalg} \citep{Verbyla2017kpcalg}, \code{dhsic.test} from package \pkg{dHSIC} \citep{pfister2017dhsic}, \code{eqdist.test} and \code{dcov.test} from package \pkg{energy} \citep{Rizzo2017energy}, \code{mdm_test} from package \pkg{EDMeasure} \citep{jin2018edmeasure}, \code{multivariance.test} from package \pkg{multivariance} \citep{bjorn2019multivariance}, \code{independence_test} from package \pkg{coin} \citep{torsten2008implementing}, \code{MINTperm} from package \pkg{IndepTest} \citep{thomas2018indeptest}, \code{hoeffD}, \code{hoeffR}, \code{pTStar} and \code{jTStar} from package \pkg{SymRC}\footnote{\url{https://github.com/Lucaweihs/SymRC}}, \code{hhg.test.k.sample} and \code{hhg.test} from package \pkg{HHG} \citep{brill2018nonparametric}, and so on. 
Among them, the functions in \pkg{stats} and \pkg{kSamples} implement the classical parametric and non-parametric hypothesis tests for 
univariate distributions and univariate random variables. 
The \pkg{TauStar} and \pkg{HellCor} packages implement two novel univariate dependence measures, 
Bergsma–Dassios sign covariance \citep{bergsma2014consistent} and Hellinger correlation \citep{geenens2018hellinger}, 
which possess admirable theoretical advantages. In short, they are designed for univariate data, 
and hence are restricted. The \pkg{Hotelling} and \pkg{FactoMineR} packages provide the multivariate extension 
of the Student's \(t\)-test and the Pearson correlation test, but the normality assumption for multivariate 
data is usually difficult to validate. The functions in \pkg{kernlab}, \pkg{kpcalg}, 
\pkg{dHSIC}, \pkg{energy}, \pkg{EDMeasure}, and \pkg{multivariance} are capable of distinguishing 
distributions and examining (mutual) independence assumptions for univariate/multivariate 
continuous/discrete data. Unfortunately, since these packages rely on energy distance \citep{szekely2004testing} 
and distance covariance \citep{szekely2007measuring} or maximum mean discrepancy \citep{arthur2012kernel} 
and Hilbert–Schmidt independence criterion \citep{gretton2005measuring}, they will totally lose power when the metric is not of strong negative type \citep{lyons2013distance} or 
the kernel is not of positive definiteness \citep{sejdinovic2013}. 
As for the \pkg{coin} package, it offers user-friendly and highly flexible interfaces for performing 
the $K$-sample and independence permutation tests on Euclidean geometry datasets under the 
framework proposed by \citet{strasser1999asymptotic}. 
The functions in \pkg{IndepTest} use mutual information, a well-known dependence measure, 
to perform the independence test between two Euclidean vectors \citep{berrett2017nonparametric}. 
The \pkg{SymRC} package provides a competitive dependence measure, symmetric rank covariances, 
to test the multivariate independence in Euclidean space \citep{drton2018symmetric}.
The functions in \pkg{HHG}, based on the work of \citet{heller2013consistent}, are known to be useful in detecting distinctions among multivariate distributions and associations between multivariate random variables in Euclidean space.  

Recently, two novel concepts, Ball Divergence \citep{pan2018ball} and Ball Covariance \citep{pan2018bcov}, 
are proposed to measure the discrepancy between two distributions and the dependence between two random objects 
in metric spaces, respectively. 
Ball Divergence (BD) enjoys a remarkable property, homogeneity-zero equivalence, and Ball Covariance (BCOV) holds another brilliant property, independence-zero equivalence.
The BD and BCOV statistics, as the empirical versions of BD and BCOV, 
can tackle the two-sample test and test of independence problems, which are special cases of the $K$-sample test and test of mutual independence problems.
The BD and BCOV statistics are both robust rank statistics, and the test procedures based on them are consistent against any general alternative hypothesis without distribution or moment assumptions \citep{pan2018ball, pan2018bcov}.
Besides, the BD statistic is proved to cope well with imbalanced data, and the BCOV statistic can be standardized to 
the Ball Correlation statistic to extract important features from ultra-high dimensional data \citep{pan2018generic}.

In this paper, we introduce a user-friendly \proglang{R} package \pkg{Ball} \shortcites{wang2018ball}\citep{wang2018ball}. The \pkg{Ball} package contributes to the open-source statistical software community in the following aspects: (i) It provides the BD two-sample test and the BCOV independence test to \proglang{R} users; (ii) It implements several dedicated design algorithms to accelerate the BD two-sample test and the BCOV independence test; (iii) It provides three powerful \(K\)-sample BD test statistics and an efficient $K$-sample permutation test procedure to distinguish distributions in metric spaces; (iv) It extends the BCOV test statistic to detect the mutual dependence among complex random objects in metric spaces; (v) It supports several generic sure independence screening procedures which are capable of extracting important features associated with complex objects in metric spaces. At present, the \pkg{Ball} package is available from the Comprehensive \proglang{R} Archive Network (CRAN) at \url{https://CRAN.R-project.org/package=Ball}.

The remaining sections are organized as follows. In Section \ref{ball-statistics}, we propose our Ball test statistics and Ball test procedures to tackle the $K$-sample test and test of mutual independence problems in metric spaces. In Section \ref{algorithm}, we introduce several novel and efficient algorithms for the Ball test statistics and Ball test procedures. Section \ref{the-ball-package} gives a detailed account for the main functions in the \pkg{Ball} package and provides two real data examples to demonstrate their usages for complex dataset. Section \ref{numerical-study} discusses the numerical performance of the Ball test statistics in the \(K\)-sample test and the test of mutual independence problems. Finally, the paper concludes with a short summary in Section \ref{conclusion}.

\section{Ball test statistics}\label{ball-statistics}

In this section, we define and illustrate the Ball Divergence (BD) and Ball Covariance (BCOV)
statistics in Section \ref{k-sample-problem-and-ball-divergence-statistic} and \ref{test-of-independence-problem-and-ball-covariance-statistic}, 
respectively. We describe the details of the Ball test procedures in Section \ref{ball-permutation-test-procedure}. 

\subsection[K-sample test and Ball Divergence statistic]{$K$-sample test and Ball Divergence statistic}\label{k-sample-problem-and-ball-divergence-statistic}

In a metric space \((V, d)\), given \(K\) independent observed samples $\mathcal{X}_{k} = \{X_{ki} | i=1,\ldots, n_k\}$, $k=1,\ldots, K$, assuming that in the $k$-th sample, $X_{k1}, \ldots, X_{kn_{k}}$ are \emph{i.i.d.} and associated with the Borel probability measure $\mu_k$. The null hypothesis of the \(K\)-sample test problem is formulated as
\begin{equation*}\label{ksamp_b}
H_{0}: \mu_{1} = \cdots = \mu_{K}.
\end{equation*}

We first revisit the BD statistic designed for the two-sample test problem \citep{pan2018ball}, a special case of the \(K\)-sample test problem, where $K=2$ and $N=n_1+n_2$. Let \(I(\cdot)\) be an indicator function. For any \(x, y, z \in V\), denote \(B(x, y)\) as a closed ball with center \(x\) and radius \(d(x, y)\), and \(\delta(x, y, z)=I(z \in B(x, y))\). Therefore, \(\delta(x, y, z)\) takes the value of 1 when \(z\) is inside the closed ball \(B(x, y)\), or 0 otherwise. Let
\begin{align*} 
P_{ij}^{\mu_1 \mu_1}=\frac{1}{n_1}\sum_{t=1}^{n_{1}}{\delta(X_{1i}, X_{1j},X_{1t})}, \; P_{ij}^{\mu_1 \mu_2}=\frac{1}{n_2}\sum_{t=1}^{n_{2}}{\delta(X_{1i}, X_{1j},X_{2t})}, \\
P_{kl}^{\mu_2\mu_1}=\frac{1}{n_1}\sum_{t=1}^{n_{1}}{\delta(X_{2k}, X_{2l},X_{1t})}, \; P_{kl}^{\mu_2\mu_2}=\frac{1}{n_2}\sum_{t=1}^{n_{2}}{\delta(X_{2k}, X_{2l},X_{2t})},
\end{align*}
then the two-sample BD statistic is defined as
\[\stbd(\mu_1, \mu_2) = \frac{1}{n_1^{2}}\sum_{i,j=1}^{n_1}{(P_{ij}^{\mu_1 \mu_1}-P_{ij}^{\mu_1 \mu_2})^{2}} + \frac{1}{n_2^{2}}\sum_{k,l=1}^{n_2}{(P_{kl}^{\mu_2 \mu_1}-P_{kl}^{\mu_2 \mu_2})^{2}}.\]
Intuitively, if $\mathcal{X}_{1}$ and $\mathcal{X}_{2}$ come from the same distribution, the proportions of the elements of $\mathcal{X}_{1}$ and $\mathcal{X}_{2}$ in the closed balls \(B(X_{1i}, X_{1j})\) and \(B(X_{2k}, X_{2l})\) are almost the same, in other words,
\(P^{\mu_{1}\mu_{1}}_{ij} \approx P^{\mu_{1}\mu_{2}}_{ij}, P^{\mu_2 \mu_1}_{kl} \approx P^{\mu_2 \mu_2}_{kl}\).
Consequently, \(\stbd\) approaches to zero in this scenario. 
Otherwise, if $\mathcal{X}_{1}$ and $\mathcal{X}_{2}$ come from two distinct distributions, then $\stbd$ is, relatively, far away from zero.

Generally, for $K > 2$ and $N=\sum_{k=1}^{K}{n_{k}}$, the definition of the \(K\)-sample BD statistic
could be to directly sum up all of the two-sample BD statistics 
\begin{align*}
\sksbd(\mu_{1}, \ldots, \mu_{K}) = \sum_{1 \leq s < t \leq K}{\bd_{n_{s}+ n_{t}}(\mu_{s}, \mu_{t})},
\end{align*}
or to find one group with the largest difference to other groups
\begin{align*}
\skmsbd(\mu_{1}, \ldots, \mu_{K}) = \max_{t}{\sum_{s=1, s \neq t}^{K}{\bd_{n_{s}+ n_{t}}{(\mu_{s}, \mu_{t})}}},
\end{align*}
or to aggregate the \(K-1\) most significant two-sample BD statistics
\begin{align*}
\skmbd(\mu_{1}, \ldots, \mu_{K}) = \sum_{k=1}^{K-1}{\bd_{(k)}},
\end{align*}
where \(\bd_{(1)}, \ldots, \bd_{(K-1)}\) are the largest \(K-1\)
two-sample BD statistics among the set 
\(\{\bd_{n_s + n_t}(\mu_s, \mu_t)| 1 \leq s < t \leq K\}.\) 
When \(K=2\), \(\sksbd, \skmbd\), and \(\skmsbd\) degenerate into \(\stbd\).

\subsection{Test of mutual independence and Ball Covariance statistic}\label{test-of-independence-problem-and-ball-covariance-statistic}
Assume that $(V_1, d_1), \ldots, (V_K, d_K)$ are metric spaces, $D$ is the Euclidean norm on $R^{K}$, and $(V, d)$ is the product metric space of $(V_1, d_1), \ldots, (V_K, d_K)$, where the product metric is defined by $d((x_1,\ldots, x_K), (y_1,\ldots, y_K))=D(d_1(x_1,y_1), \ldots, d_K(x_K,y_K))$. 
Suppose $\mathcal{X} = \{(X_{i1},\ldots, X_{iK})| i=1, \ldots, N\}$ is a sample with $N$ \emph{i.i.d.} observations in the product metric space $V$, whose associated joint Borel probability measure and marginal Borel probability measures are $\mu$ and $\mu_1, \ldots, \mu_K$. The null hypothesis of the test of mutual independence problem is formulated as
\begin{align*}\label{ind_b}
H_{0}: \mu=\mu_1 \otimes \cdots \otimes  \mu_K.
\end{align*}
For \(x, y, z \in V_i\), denote \(B_i(x, y)\) as a closed ball with center \(x\) and radius \(d_i(x, y)\), and
\(\delta_i(x, y, z)=I(z \in B_i(x, y))\). Thus, \(\delta_i(x, y, z)\) is an indicator taking the value of 1 if \(z\) is
within the closed ball \(B_i(x, y)\), or 0 otherwise. Let 
\begin{equation*}
P_{ij}^{\mu}=\frac{1}{N}\sum_{t=1}^{N}\prod_{k=1}^K{\delta_k(X_{ik}, X_{jk}, X_{tk})},
\end{equation*}
and
\begin{equation*}
P_{ij}^{\mu_k}=\frac{1}{N}\sum_{t=1}^{N}{\delta_k(X_{ik}, X_{jk}, X_{tk}),} \ \ k=1,\ldots, K,
\end{equation*}
then our BCOV statistic can be presented as
\begin{equation*}\label{BCOVN}
  \scbcov(\mu_1, \ldots, \mu_K)=\frac{1}{N^{2}}\sum_{i,j=1}^{N}{(P_{ij}^{\mu}-\prod_{k=1}^KP_{ij}^{\mu_k})^{2}}.
\end{equation*}
We provide a heuristic explanation for $\scbcov$. If $\mu = \mu_1 \otimes \cdots \otimes  \mu_K$, the proportion of the elements of $\mathcal{X}$ in $\otimes_{k=1}^K B_k(X_{ik}, X_{jk})$ should be close to the product of the proportions of $X_{1k}, \ldots, X_{Nk}$ in \(B_k(X_{ik}, X_{jk})\), i.e., $P_{ij}^{\mu} \approx \prod_{k=1}^KP_{ij}^{\mu_k}$.
Since \(\mbox{BCov}_N^{K}\) is the average of \(\{(P_{ij}^{\mu}-\prod_{k=1}^KP_{ij}^{\mu_k})^{2}|i, j = 1, \ldots, N\}\),
a significantly large \(\scbcov\) implies that the null hypothesis
\(\mu=\mu_1 \otimes \cdots \otimes  \mu_K\) is implausible.

The BCOV statistic can be extended with positive weights $\hat{\omega}_k(X_{ik}, X_{jk}), k=1, \ldots, K$, 
\begin{align*}
  \sbcov(\mu_1, \ldots, \mu_K)=\frac{1}{N^{2}}\sum_{i,j=1}^{N}{(P_{ij}^{\mu}-\prod_{k=1}^K P_{ij}^{\mu_k})^{2}\prod_{k=1}^K \hat{\omega}_k(X_{ik}, X_{jk})}.
\end{align*}
As a more general dependence measure framework based on the BCOV statistic, 
such a weighted extension not only allows flexible test statistics but also connects the BCOV statistic with HHG. 
Several choices for the weights are feasible. For example, we could choose the probability weight \(\hat{\omega}_k(X_{ik}, X_{jk}) = [P_{ij}^{\mu_k}]^{-1}\) and the Chi-square weight \(\hat{\omega}_k(X_{ik}, X_{jk})= [P_{ij}^{\mu_k}(1-P_{ij}^{\mu_k})]^{-1}\), and denote their corresponding statistics as \(\spbcov\) and \(\schibcov\), respectively. 
\(\spbcov\) focuses on smaller balls, while \(\schibcov\) standardizes \((P_{ij}^{\mu}-\prod_{k=1}^K P_{ij}^{\mu_k})^{2}\) by the variance of \(\delta_k(X_{ik}, X_{jk}, X_{tk})\) (\(i, j\) are fixed). Furthermore, \(\schibcov (K=2)\) is asymptotically equivalent to HHG \citep{pan2018bcov}. 
From the definitions of $\skcbcov, \skpbcov$, and $\skchibcov$, 
we may expect: 
(i) $\skcbcov$ is good at detecting linear relationships, especially when noises are influential, 
because treating each ball indiscriminately is a reasonable strategy which keeps weights away from potential instabilities; 
(ii) $\skpbcov$ has more power in detecting strong nonlinear relationships 
since paying more attention to smaller balls makes $\skpbcov$ tending to detect the locally linear relationship;
(iii) $\skchibcov$ (or HHG) is an intermediate between $\skcbcov$ and $\skpbcov$.

The BCOV statistic can be normalized. Let 
\begin{align*}
\sbcov(\mu_k)=\frac{1}{N^{2}}\sum_{i,j=1}^{N}{[P_{ij}^{\mu_k}-(P_{ij}^{\mu_k})^{K}]^{2}}(\hat{\omega}_k(X_{ik}, X_{jk}))^{K}, \ \ k= 1, \ldots, K,
\end{align*}
then the normalized version of the BCOV statistic is defined as the square root of
\begin{align*}
\skbcor(\mu_1, \ldots, \mu_K) = \sbcov(\mu_1, \ldots, \mu_K)/\sqrt{\prod_{k=1}^{K}\sbcov(\mu_k)},
\end{align*}
if $\prod_{k=1}^{K}\sbcov(\mu_k) > 0$, or 0 otherwise. $\skbcor(K=2)$ is the Ball Correlation statistic \citep{pan2018bcov} which ranges from 0 to 1.

\subsection{Ball permutation test procedure}\label{ball-permutation-test-procedure}
After computing an observed Ball test statistic, say $B$, the permutation methodology described in \citet{efron1994introduction} and \citet{davison1997bootstrap} is employed to derive the $p$ value of the Ball test procedures in a distribution-free manner. We specify the permutation procedures for the $K$-sample test and test of mutual independence problems below.

For the $K$-sample test problem, let $\textbf{k}_{n_k}$ be an $n_k$-dimensional vector whose entries are all $k$, and let $L = (\textbf{1}^{\top}_{n_1}, \textbf{2}^{\top}_{n_2}, \ldots, \textbf{K}^{\top}_{n_K})^{\top}$ be the group label vector of the pooled sample $\mathcal{X} = \mathcal{X}_{1} \cup \cdots \cup \mathcal{X}_{K}$. Denote $L^{*}$ as a permutation of $L$ and $\{Y_i\}_{i=1}^{N}$ as $\mathcal{X}$, the shuffled pooled sample associated with $L^{*}$ is $\mathcal{X}^{*} = \mathcal{X}^{*}_{1} \cup \cdots \cup \mathcal{X}^{*}_{K}$, where $\mathcal{X}^{*}_{k} = \{ Y_i \in \mathcal{X}| L^{*}_i = k, i = 1, \ldots, N \}$. With the shuffled pooled sample $\mathcal{X}^{*}$, we can compute the $K$-sample BD statistic. The permutation is replicated $M$ times to derive the $K$-sample BD statistics under the null hypothesis: $B^{\prime}_{1}, \ldots, B^{\prime}_{M}$. Finally, the $p$ value is computed according to:
\begin{equation}\label{p-value}
p \text{ value} = \frac{1 + \sum_{m=1}^{M} I(B_m^{\prime} \geq B)}{1+M}.
\end{equation}
With respect to the test of mutual independence problem, each permutation is performed following this manner: for each $k \in \{1, \ldots, K\}$, $X_{1k}, \ldots, X_{Nk}$ are randomly shuffled while $X_{1k^{\prime}}, \ldots, X_{Nk^{\prime}}$ $(k^{\prime} \neq k)$ are fixed. The permutation is replicated $M$ times and the $p$ value is estimated by Equation~\ref{p-value}.

\section{Algorithm}\label{algorithm}
The computational complexities of the Ball test statistics are \(O(N^3)\) if we compute them according to their definitions; however, in many cases, their computational complexities can reduce to a more moderate level (see Table~\ref{computation_complexity}) by efficient algorithms utilizing the rank nature of the Ball test statistics. The rank nature motivates the acceleration in three aspects. 
First, the computation procedure of the Ball test statistics can avoid repeatedly evaluating whether a point is in a closed ball 
because \(n_{t}P_{ij}^{\mu_{s}\mu_{t}} (s, t \in \{1, \ldots, K\})\) and \(NP^{\mu_k}_{ij} (k=1, \ldots, K)\) are related to some ranks. 
Second, for the $K$-sample permutation test, performing ranking procedures could be more efficient by preserving information for the first time when we compute the $K$-sample BD statistics. Third, for univariate datasets, we can optimize the ranking procedure for computing \(n_{t}P_{ij}^{\mu_{s}\mu_{t}} (s, t \in \{1, \ldots, K\})\) and \(NP^{\mu_k}_{ij} (k=1, \ldots, K)\) without preserving auxiliary information. The three aspects will be illustrated in Section \ref{rank-based-algorithm}, \ref{optimized-permutation-test-procedure}, and \ref{fast-algorithm-for-univariate-data}, respectively. 
For simplicity, we assume there is no ties among datasets. For our \pkg{Ball} package, 
it can properly handle tied data and compute the exact Ball test statistics.

Unfortunately, in the case of measuring mutual dependence among at least three random objects, it is not easy to optimize the computation procedure of the BCOV statistic. Nevertheless, with engineering optimizations such as multi-threading, the time consumption of computing the BCOV statistic can be cut down.

\begin{table}[ht]
  \renewcommand\arraystretch{1.3}
  \renewcommand\tabcolsep{10.0pt}
  {\large
  \begin{center}
  \begin{tabular}{l|cc}
  \toprule
  Statistics &  Univariate & Other \\
  \hline
  $\sksbd / \skmsbd / \skmbd$ & $O(N^2)$ & $O(N^2)$ \\
  \hline
  $\sbcov (K = 2) $ & $O(N^2)$ & $O(N^2 \log{N})$ \\
  \bottomrule
  \end{tabular}
  \vspace*{-10pt}
  \end{center}}
  \vspace*{-5pt}
  \caption{The optimized computational complexities of several Ball test statistics.}\label{computation_complexity}
  \vspace*{-5pt}
\end{table}

\subsection{Rank based algorithm}\label{rank-based-algorithm}
We first recap the \(O(N^2\log{N})\) algorithm for \(\stbd\) proposed by
\citet{pan2018ball}. Assume the pairwise distance matrix of
\(\mathcal{X} = \mathcal{X}_1 \cup \mathcal{X}_2\) is:
\begin{align*}
 D^{\mathcal{XX}}=
 \begin{pmatrix}
D^{\mathcal{X}_{1}\mathcal{X}_{1}} & D^{\mathcal{X}_{1}\mathcal{X}_{2}} \\
D^{\mathcal{X}_{2}\mathcal{X}_{1}} & D^{\mathcal{X}_{2}\mathcal{X}_{2}}
\end{pmatrix}_{N \times N}.
\end{align*}
Notice that, $n_1P^{\mu_{1}\mu_{1}}_{ij}$ is the rank of $d(X_{1i}, X_{1j})$ among 
the $i$-th row of $D^{\mathcal{X}_{1}\mathcal{X}_{1}}$, and $n_1P^{\mu_{1}\mu_{1}}_{ij} + n_2P^{\mu_{1}\mu_{2}}_{ij}$ 
is the rank of $d(X_{1i}, X_{1j})$ among the $i$-th row of $D^{\mathcal{X}\mathcal{X}}$.
Consequently, after spending \(O(N^2\log{N})\) time on ranking \(D^{\mathcal{X}_{1} \mathcal{X}_{1}}\) and \(D^{\mathcal{XX}}\) row by row to obtain their corresponding rank matrices \(R^{\mathcal{X}_{1}\mathcal{X}_{1}}\) and \(R^{\mathcal{XX}}\), we only need \(O(1)\) time to compute \(n_1P^{\mu_{1}\mu_{1}}_{ij}\) and \(n_1P^{\mu_{1}\mu_{1}}_{ij} + n_2P^{\mu_{1}\mu_{2}}_{ij}\) by directly extracting the \((i, j)\)-elements of \(R^{\mathcal{X}_{1}\mathcal{X}_{1}}\) and \(R^{\mathcal{XX}}\). 
Therefore, computing $\{ (P^{\mu_{1}\mu_{1}}_{ij}, P^{\mu_{1}\mu_{2}}_{ij}) | i, j = 1, \ldots, n_{1} \}$ is of $O(N^2 \log{N})$ time complexity, and similarly for computing $\{(P^{\mu_{2}\mu_{1}}_{kl}, P^{\mu_{2}\mu_{2}}_{kl}) | k, l = 1, \ldots, n_{2})\}$. 
In summary, for $\stbd$ and the $K$-sample BD statistics, their time complexities are $O(N^2 \log{N})$.

With respect to \(\sbcov (K=2)\), the aforementioned algorithm can be directly applied to calculate \(P_{ij}^{\mu_{1}}\) and \(P_{ij}^{\mu_{2}}\)
within \(O(N^2\log{N})\) time. To compute \(\{NP^{\mu}_{ij}|i, j = 1, \ldots, N\}\) within \(O(N^2 \log{N})\) time, we first rearrange
\(\{(d_1(X_{i1}, X_{j1}), d_2(X_{i2}, X_{j2})) | j = 1, \ldots, N\}\) to
\(\{(d_1(X_{i1}, X_{i_{j}1}), d_2(X_{i2}, X_{i_{j}2})) | j = 1, \ldots, N)\}\), where
\(i_j\) is the location of the \(j\)-th smallest value among
\(\{d_1(X_{i1}, X_{j1}) | j = 1, \ldots, N\}\). Further, for $j = 1, \ldots, N$, we have
\begin{equation}\label{count_of_smaller}
\begin{aligned}
NP^{\mu}_{ii_{j}} 
&= \sum_{t=1}^{N}I(d_1(X_{i1}, X_{i_{t}1}) \leq d_1(X_{i1}, X_{i_{j}1}))I(d_2(X_{i2}, X_{i_{t}2}) \leq d_2(X_{i2}, X_{i_{j}2})) \\
&= \sum_{t=1}^{j}I(d_2(X_{i2}, X_{i_{t}2}) \leq d_2(X_{i2}, X_{i_{j}2})) \\
&= \sum_{t=1}^{N}I(d_2(X_{i2}, X_{i_{t}2}) \leq d_2(X_{i2}, X_{i_{j}2})) - \sum_{t=j + 1}^{N}I(d_2(X_{i2}, X_{i_{t}2}) \leq d_2(X_{i2}, X_{i_{j}2})) \\
&= NP_{i i_{j}}^{\mu_2} - \sum_{t=j + 1}^{N}I(d_2(X_{i2}, X_{i_{t}2}) \leq d_2(X_{i2}, X_{i_{j}2})). \\
\end{aligned}
\end{equation}
Owing to Equation~\ref{count_of_smaller}, the computation of \(\{NP^{\mu}_{ii_j}| j = 1, \ldots, N\}\) could be turned into computing \(\{\sum_{t=j + 1}^{N}I(d_2(X_{i2}, X_{i_{t}2}) \leq d_2(X_{i2}, X_{i_{j}2}))| j = 1, \ldots, N\}\). Notice that, given the array $A=\{d_2(X_{i2}, X_{i_{j}2})|j = 1, \ldots, N\}$, \(\sum_{t=j + 1}^{N}I(d_2(X_{i2}, X_{i_{t}2}) \leq d_2(X_{i2}, X_{i_{j}2}))\) is the number of the elements of $A$ which are not only behind \(d_2(X_{i2}, X_{i_{j}2})\) but also no larger than \(d_2(X_{i2}, X_{i_{j}2})\).
Therefore, computing \(\{\sum_{t=j + 1}^{N}I(d_2(X_{i2}, X_{i_{t}2}) \leq d_2(X_{i2}, X_{i_{j}2}))| j = 1, \ldots, N\}\) is the typical ``count of smaller numbers after self'' problem\footnote{\url{https://leetcode.com/problems/count-of-smaller-numbers-after-self/}}, which can be solved within \(O(N\log{N})\) time via well designed algorithms such as binary indexed tree, binary search tree, and merge sort. Our implementation (See Appendix) utilizes merge sort to resolve the problem due to its efficiency. Thus, the time complexity of $\{NP^{\mu}_{ij}| i,j = 1, \ldots, N\}$ reduces to $O(N^2 \log{N})$, and finally, the computation of $\sbcov (K=2)$ costs $O(N^2 \log{N})$ time.

\subsection[Optimized K-sample permutation test procedure]{Optimized $K$-sample permutation test procedure}\label{optimized-permutation-test-procedure}
In this section, an efficient procedure of $O(N^2)$ time complexity is introduced to compute the
$K$-sample BD statistics on a shuffled dataset $\mathcal{X}^{*}$. 
According to the definition of the $K$-sample BD statistics, to reduce their time complexity to $O(N^2)$, 
we should reduce the time complexity of any $\text{BD}_{n_s + n_t}$ to $O(N^2)$. 
From Section \ref{rank-based-algorithm}, to reduce the time complexity of $\text{BD}_{n_s + n_t}$ to $O(N^2)$,
we need to compute the row-wise ranks of pairwise distance matrices $D^{\mathcal{X}^{*}_s\mathcal{X}^{*}_s}, D^{\mathcal{X}^{*}_t\mathcal{X}^{*}_t},$ and $D^{\mathcal{X}^{*}_{s,t}\mathcal{X}^{*}_{s,t}} (\mathcal{X}^{*}_{s,t} = \mathcal{X}^{*}_s \cup \mathcal{X}^{*}_t)$ within $O(N^2)$ time.
We propose Algorithm \ref{permuted_bd} to achieve this goal. 
The core of Algorithm \ref{permuted_bd} is utilizing an $N \times N$ order information matrix $I^{\mathcal{XX}}$ as well as 
an $N$-dimensional vector $G$. 
For the order information matrix $I^{\mathcal{XX}}$, \(I^{\mathcal{XX}}_{ij}=q\) means that the \(j\)-th smallest element of the \(i\)-th row of pairwise distance matrix \(D^{\mathcal{XX}}\) is \(D^{\mathcal{XX}}_{iq}\). 
$I^{\mathcal{XX}}$ can be obtained for the first time when we rank each row of the pairwise distance matrix $D^{\mathcal{XX}}$. 
Concerning the vector $G$, it is related to the shuffled group label vector $L^{*}$ and 
the cumulative sample size vector $C = (0, n_1, \ldots, \sum_{k=1}^{K-1}n_k)$. 
Specifically, when $L^{*}_i = k$, $G_{i} = C_{k} + \sum_{j=1}^{i-1}I(L_j^{*} = k) + 1$ if $i > 1$, or $G_{i} = C_{k} + 1$ if $i = 1$. 
By scanning $I^{\mathcal{XX}}$ in a row-wise manner, Algorithm \ref{permuted_bd} can assign a proper rank value
to the element of the row-wise rank matrices of $D^{\mathcal{X}^{*}_s\mathcal{X}^{*}_s}, D^{\mathcal{X}^{*}_t\mathcal{X}^{*}_t},$ 
and $D^{\mathcal{X}^{*}_{s,t}\mathcal{X}^{*}_{s,t}}$ with the help of the vector $G$.
\begin{algorithm}[ht]
\caption{Optimized algorithm for computing the row-wise rank matrices of $D^{\mathcal{X}_{s}^{*}\mathcal{X}_{s}^{*}}, D^{\mathcal{X}_{t}^{*}\mathcal{X}_{t}^{*}}$, and $D^{\mathcal{X}_{s, t}^{*}\mathcal{X}_{s, t}^{*}}$}
\label{permuted_bd}
\begin{algorithmic}[1]
\REQUIRE The sample size of the $s$-th group $n_s$, the cumulative sample size vector $C$, the order information matrix $I^{\mathcal{XX}}$ as well as the $N$-dimensional vector $G$.
\STATE Initialize all element of the row-wise rank matrices $R^{\mathcal{X}^{*}_s\mathcal{X}^{*}_s}$/$R^{\mathcal{X}^{*}_t\mathcal{X}^{*}_t}$/$R^{\mathcal{X}^{*}_{s,t}\mathcal{X}^{*}_{s,t}}$ of $D^{\mathcal{X}^{*}_s\mathcal{X}^{*}_s}$/$D^{\mathcal{X}^{*}_t\mathcal{X}^{*}_t}$/$D^{\mathcal{X}^{*}_{s,t}\mathcal{X}^{*}_{s,t}}$ with 0.
\FOR {$i = 1, \ldots, N$}
  \STATE $g \leftarrow L^{*}_{i}.$
	  \IF {$g = s$ \OR $g = t$}
	  \STATE $rank_1 \leftarrow 1, rank_2 \leftarrow 1, rank_{3} \leftarrow 1$.
	  \FOR {$j = 1, \ldots, N$}
	    \STATE $o \leftarrow I^{\mathcal{X}\mathcal{X}}_{i, j}$, $g^\prime \leftarrow L^{*}_{o}$.
	    \IF {$g = g^\prime$}
	      \IF {$g = s$}
	        \STATE $r \leftarrow G_i - C_s, c \leftarrow G_o - C_s$,
	        \STATE $R^{\mathcal{X}^{*}_{s}\mathcal{X}^{*}_{s}}_{rc} \leftarrow rank_1, R^{\mathcal{X}^{*}_{s, t}\mathcal{X}^{*}_{s, t}}_{rc} \leftarrow rank_{3}$, 
	        \STATE $rank_1 \leftarrow rank_1 + 1$.
	      \ELSIF {$g = t$}
	        \STATE $r \leftarrow G_i - C_t, c \leftarrow G_o - C_t$, 
          \STATE $r^{\prime} \leftarrow G_i - C_t + n_s, c^{\prime} \leftarrow G_o - C_t + n_s$,
	        \STATE $R^{\mathcal{X}^{*}_{t}\mathcal{X}^{*}_{t}}_{rc} \leftarrow rank_2, R^{\mathcal{X}^{*}_{s, t}\mathcal{X}^{*}_{s, t}}_{r^{\prime}c^{\prime}} \leftarrow rank_{3}$,
	        \STATE $rank_2 \leftarrow rank_2 + 1$. 
	      \ENDIF
	    \ELSE
	      \IF {$g = s$ \AND $g^\prime = t$}
	        \STATE $r \leftarrow G_i - C_s, c \leftarrow G_o - C_t + n_s$.
	      \ELSIF {$g = t$ \AND $g^\prime = s$}
	        \STATE $r \leftarrow G_i - C_t + n_s, c \leftarrow G_o - C_s$.
	      \ENDIF
        \STATE $R^{\mathcal{X}^{*}_{s, t}\mathcal{X}^{*}_{s, t}}_{rc} \leftarrow rank_{3}$.
	    \ENDIF
      \STATE $rank_{3} \leftarrow rank_{3} + 1$.
	  \ENDFOR
  \ENDIF
\ENDFOR
\RETURN $R^{\mathcal{X}_{s}^{*}\mathcal{X}_{s}^{*}}, R^{\mathcal{X}_{t}^{*}\mathcal{X}_{t}^{*}},$ and $R^{\mathcal{X}_{s, t}^{*}\mathcal{X}_{s, t}^{*}}$.
\end{algorithmic}
\end{algorithm}

\subsection{Fast algorithm for univariate data}\label{fast-algorithm-for-univariate-data}
We first consider \(\stbd\). For convenience, assume $X_{11}, \ldots, X_{1n_{1}}$ have been sorted in ascending order. For univariate data, we have
\begin{equation}\label{univariate_ball}
\begin{aligned}
n_1P^{\mu_1\mu_1}_{ij} &= \sum_{t=1}^{n_1}{I(|X_{1t} - X_{1i}| \leq |X_{1i} - X_{1j}|)} \\
&= \sum_{t=1}^{n_1}{I(X_{1i} - |X_{1i} - X_{1j}| \leq X_{1t} \leq X_{1i} + |X_{1i} - X_{1j}|)}.
\end{aligned}
\end{equation}
Let \(X_{1l_{ij}}\) be the smallest value satisfying \(X_{1l_{ij}} \geq X_{1i} - |X_{1i} - X_{1j}|\) and \(X_{1r_{ij}}\) be the largest element satisfying \(X_{1r_{ij}} \leq X_{1i} + |X_{1i} - X_{1j}|\). Thanks to Equation~\ref{univariate_ball}, it is easy to verify that \(n_1P^{\mu_1\mu_1}_{ij} = r_{ij} - l_{ij} + 1\), and consequently, an alternative way to compute $n_1P^{\mu_1\mu_1}_{ij}$ is to find out $X_{1l_{ij}}$ and $X_{1r_{ij}}.$ Inspired by this, we develop Algorithm \ref{bd_same_group} to accomplish the computation of \(\{n_1P_{ij}^{\mu_1\mu_1} | j = 1,\ldots, n_1\}\) in linear time. Through slightly modifying Algorithm~\ref{bd_same_group}, the computational complexity of \(\{n_1P^{\mu_1\mu_1}_{ij} + n_2P^{\mu_1\mu_2}_{ij}| j=1, \ldots, n_{1}\}\) also reduces to \(O(N)\), and hence, the computational complexity of \(\{(n_1P^{\mu_1\mu_1}_{ij}, n_1P^{\mu_1\mu_2}_{ij}) | i, j = 1,\ldots, n_1\}\) reduces to \(O(N^2)\). Similarly, the time complexity of \(\{(n_1P^{\mu_2\mu_1}_{kl}, n_2P^{\mu_2\mu_2}_{kl}) | i,j=1,\ldots,n_2\}\) is \(O(N^2)\). In summary, the computational complexity of \(\stbd\) is \(O(N^2)\) for univariate distributions, so are $\sksbd, \skmsbd$, and $\skmbd$. 
\begin{algorithm}[ht]
\caption{Fast algorithm for $\{n_1P_{ij}^{\mu_1\mu_1}| j = 1,\ldots, n_1\}$ in the univariate case}
\label{bd_same_group}
\begin{algorithmic}[1]
\REQUIRE A sorted array $\{X_{11}, \ldots, X_{1 n_1}\}$ with ascending order.
\STATE Initialize $l=1, r=n_1$.
\WHILE {$l \leq r$}
    \IF {$X_{1r}-X_{1i} \geq X_{1i}-X_{1l}$}
        \STATE $n_1P_{ir}^{\mu_1\mu_1} \leftarrow r - l + 1$,
        \STATE $r \leftarrow r - 1$.
    \ELSE
        \STATE $n_1P_{il}^{\mu_1\mu_1} \leftarrow r - l + 1$,
        \STATE $l \leftarrow l + 1$.
    \ENDIF
\ENDWHILE
\RETURN $\{n_1P_{ij}^{\mu_1\mu_1}| j = 1,\ldots, n_1\}$
\end{algorithmic}
\end{algorithm}

As for \(\sbcov (K=2) \), by slightly modifying Algorithm \(\ref{bd_same_group}\), the time complexity of computing \(\{(NP_{ij}^{\mu_1}, NP_{ij}^{\mu_2}) | i, j=1,\ldots, N\}\) reduces to \(O(N^2)\) when \(X_{i1}, X_{i2}\) are univariate. Further, retaining \((l^{1}_{ij}, r^{1}_{ij})\) and \((l^{2}_{ij}, r^{2}_{ij})\) which satisfy \(NP_{ij}^{\mu_1} = r^{1}_{ij} - l^{1}_{ij} + 1\) and \(NP_{ij}^{\mu_2} = r^{2}_{ij} - l^{2}_{ij} + 1\), we can accomplish the computation of \(\{P^{\mu}_{ij}|i,j=1, \ldots, N\}\) within quadratic time by Algorithm~\ref{univariate_bcov}. 
The key of Algorithm~\ref{univariate_bcov} is employing the inclusion–exclusion principle which is also used in \citet{heller2016consistent}.
In summary, with Algorithms~\ref{bd_same_group} and \ref{univariate_bcov}, the time complexity of \(\sbcov (K=2)\) is $O(N^2)$.
\begin{algorithm}[ht]
\caption{Fast algorithm for $\{P^{\mu}_{ij}|i,j=1,\ldots,N\}$ in the univariate case}
\label{univariate_bcov}
\begin{algorithmic}[1]
\REQUIRE A set including $N$ bivariate observations $\{ (X_{i1}, X_{i2})| i=1, \ldots, N \}$; a set containing lower and upper bound pairs $\{(l^{1}_{ij}, r^{1}_{ij}),(l^{2}_{ij}, r^{2}_{ij})| i, j = 1, \ldots, N \}$.
\STATE Compute the rank of $\{ X_{11}, \ldots, X_{N1} \}$ and $\{ X_{12}, \ldots, X_{N2} \}$, respectively, and denote them as $\{ r^{1}_{1}, \ldots, r^{1}_{N} \}$ and $\{ r^{2}_{1}, \ldots, r^{2}_{N} \}$.
\STATE Pay $O(N^2)$ time to compute the values of bivariate empirical cumulative distribution 
$F_{N}(x, y)=\frac{1}{N}\sum_{t=1}^{N}{I(r^{1}_{t} \leq x, r^{2}_{t} \leq y)}$   
on $\{(i, j)|i, j = 1, \ldots, N\}$, and set $F_{N}(0, y) = F_{N}(x, 0) = 0$.
\FOR {$i = 1, \ldots, N $}
  \FOR {$j = 1, \ldots, N $}
    \STATE $P_{ij}^{\mu} \leftarrow F_{N}(r^{1}_{ij}, r^{2}_{ij}) - F_{N}(l^{1}_{ij}-1, r^{2}_{ij}) - F_{N}(r^{1}_{ij}, l^{2}_{ij}-1) + F_{N}(l^{1}_{ij}-1, l^{2}_{ij}-1).$
  \ENDFOR
\ENDFOR
\RETURN $\{P^{\mu}_{ij}|i,j=1,\ldots,N\}$
\end{algorithmic}
\end{algorithm}

\section{The Ball package}\label{the-ball-package}
In this section, we introduce the \pkg{Ball} package, an \proglang{R} package which implements the Ball test statistics and procedures introduced in Section
\ref{ball-statistics} as well as the algorithms illustrated in Section \ref{algorithm}. The core of \pkg{Ball} is programmed in \proglang{C} to
improve computational efficiency. Moreover, we employ the parallel computing technique in the Ball test procedures to speed up the computation. To be specific, during the permutation procedure, multiple Ball test statistics are concurrently calculated with \proglang{OpenMP} which supports the multi-platform shared memory multiprocessing programming in \proglang{C} level. 
Aside from speed, the \pkg{Ball} package is concise and user-friendly. An \proglang{R} user can conduct the \(K\)-sample and independence tests via \code{bd.test} and \code{bcov.test} functions in the \pkg{Ball} package, respectively. 

We supply instructions and primary usages for \code{bd.test} and \code{bcov.test} functions in Section~\ref{function-bd.test-and-bcov.test}. Additionally, in Section \ref{examples}, two real data examples are provided to demonstrate how to use these functions to tackle data drawn from manifold spaces. 

\subsection[Functions bd.test and bcov.test]{Functions \code{bd.test} and \code{bcov.test}}\label{function-bd.test-and-bcov.test}

The functions \code{bd.test} and \code{bcov.test} are programmed for the \(K\)-sample test and test of mutual independence problems, respectively. The default usages of two functions are:
\begin{CodeInput}
bd.test.default(x, y = NULL, num.permutations = 99, distance = FALSE, 
  size = NULL, seed = 1, num.threads = 0, kbd.type = "sum", ...)
bcov.test.default(x, y = NULL, num.permutations = 99, distance = FALSE, 
  weight = FALSE, seed = 1, num.threads = 0, ...)
\end{CodeInput}
The arguments of the two functions are described as follows.
\begin{itemize}
\item{}{\code{x}: a numeric vector, matrix, or data.frame, or a list containing at least two numeric vectors, matrices, or data.frames.}
\item{}{\code{y}: a numeric vector, matrix, or data.frame.}
\item{}{\code{num.permutations}: the number of permutation replications, must be a non-negative integer. Default: \code{num.permutations = 99}.}
\item{}{\code{distance}: if \code{distance = TRUE}, \code{x} is considered as a distance matrix, or a list containing distance matrices in the test of mutual independence problem. And \code{y} is considered as a distance matrix only in the test of independence problem. Default: \code{distance = FALSE}.}
\item{}{\code{size}: a vector recording the sample size of $K$ groups. It is only available for \code{bd.test}.}
\item{}{\code{weight}: a logical value or character string used to choose the form of $\hat{\omega}_{k}(X_{ik}, X_{jk})$. 
If \code{weight = FALSE} or \code{weight = "constant"}, the result of $\scbcov$ based test is displayed. 
Alternatively, \code{weight = TRUE} or \code{weight = "probability"} indicates the probability weight is chosen while 
setting \code{weight = "chisquare"} means selecting the Chi-square weight. 
From the definitions of $\skcbcov, \skpbcov$, and $\skchibcov$, they are quite similar and could be computed at the same time. 
Therefore, \code{bcov.test} simultaneously computes $\skcbcov, \skpbcov$, $\skchibcov$, and their corresponding $p$~values. 
Users could get other statistics and $p$~values from the \code{complete.info} element of output without re-running \code{bcov.test}. 
At present, this arguments is only available for \code{bcov.test}. Any unambiguous substring can be given. Default: \code{weight = FALSE}.}
\item{}{\code{seed}: the random seed. Default: \code{seed = 1}.}
\item{}{\code{num.threads}: the number of threads used in the Ball test procedures. If \code{num.threads = 0}, all available cores are used. Default: \code{num.threads = 0}.}
\item{}{\code{kdb.type}: a character string used to choose the $K$-sample BD statistics. 
Setting \code{kdb.type = "sum"}, \code{kdb.type = "summax"}, or \code{kdb.type = "max"}, 
we choose $\sksbd$, $\skmsbd$, or $\skmbd$ to test the equality of distributions. 
Notice that, the three $K$-sample BD statistics are very similar, since their only difference is the aggregation strategy for the two-sample BD statistics. 
Therefore, the \code{bd.test} function simultaneously computes $\sksbd, \skmsbd$, $\skmbd$, and their corresponding $p$~values.
Users could get other statistics and $p$~values from the \code{complete.info} element of output without re-running \code{bd.test}. 
This arguments is only available for the \code{bd.test} function. Any unambiguous substring can be given. Default: \code{kdb.type = "sum"}.}
\end{itemize}

If \code{num.permutations > 0}, the output is a \code{htest} class object
similar to the object returned by the \code{t.test} function. The output object contains the Ball test statistic value (\code{statistic}), 
the $p$~value of the test (\code{p.value}), the number of permutation replications
(\code{replicates}), a vector recording the sample size (\code{size}), a
\code{list} mainly containing two vectors, where the first vector is 
$\sksbd, \skmsbd$, and $\skmbd$ for \code{bd.test} or 
$\skcbcov$, $\skpbcov$, and $\skchibcov$ for \code{bcov.test}, 
and the second vector is the corresponding $p$~values of the tests (\code{complete.info}), 
a character string declaring the alternative hypothesis (\code{alternative}), 
a character string describing the hypothesis test (\code{method}), 
and a character string giving the name and helpful information of the data (\code{data.name});
if \code{num.permutations = 0}, only the Ball test statistic value is returned.

To give quick examples, we carry out the Ball test on two synthetic datasets to check whether the null hypothesis can be rejected when distributions are different or random variables are associated indeed.

For the \(K\)-sample test problem (\(K=2\)), we generate two univariate datasets \(\{ X_{1i} \}_{i=1}^{50}\) and \(\{ X_{2i} \}_{i=1}^{50}\) with different location parameters, where 
\begin{align*}
X_{1i} \sim N(0, 1), \; X_{2i} \sim N(1, 1), \; i=1, \ldots, 50.
\end{align*}
The detailed \proglang{R} code is as follows.
\begin{CodeInput}
R> library("Ball")
R> set.seed(1)
R> x <- rnorm(50)
R> y <- rnorm(50, mean = 1)
R> bd.test(x = x, y = y)
\end{CodeInput}

\begin{CodeOutput}
    2-sample Ball Divergence Test

data:  x and y
number of observations = 100, group sizes: 50 50
replicates = 99
bd = 0.092215, p-value = 0.01
alternative hypothesis: distributions of samples are distinct
\end{CodeOutput}

In this example, the \code{bd.test} function yields the \(\stbd\) value 0.0922 and the $p$~value 0.01 when the permutation replicate is 99. At the usual significance level of 0.05, we should reject the null hypothesis. Thus,
the test result is concordant to the data generation mechanism.

In regard to the test of mutual independence problem, we sample 100 \emph{i.i.d.} observations from the multivariate normal distribution to perform the mutual independence test based on \(\scbcov\). The detailed \proglang{R} code is demonstrated below.

\begin{CodeInput}
R> library("mvtnorm")
R> set.seed(1)
R> cov_mat <- matrix(0.3, 3, 3)
R> diag(cov_mat) <- 1
R> data_set <- rmvnorm(n = 100, mean = c(0, 0, 0), sigma = cov_mat)
R> data_set <- as.list(as.data.frame(data_set))
R> bcov.test(x = data_set)
\end{CodeInput}

\begin{CodeOutput}
    Ball Covariance test of mutual independence

data:  data_set
number of observations = 100
replicates = 99, weight: constant
bcov.constant = 0.00063808, p-value = 0.02
alternative hypothesis: random variables are dependent
\end{CodeOutput}

The output of \code{bcov.test} shows that \(\scbcov\) is \(6.38 \times 10^{-4}\) when the constant weight is used. 
The $p$~value is 0.02, and at the usual significance level of 0.05, we conclude that the three univariate variables are mutually dependent.

\subsection{Examples}\label{examples}

\subsubsection{Wind direction dataset}\label{two-sample-test-for-wind-data}

We consider the hourly recorded wind speed and wind direction in the Atlantic coast of Galicia in winter from 2003 until 2012, provided in the \proglang{R} package \pkg{NPCirc} \citep{npcirc2014maria}. In this dataset, there exist 19488 observations, and each observation includes six variables: day, month, year, hour, wind speed, and wind direction (in degrees). It is of interest to see whether there are any wind direction differences between the first and last weeks of 2007-08 winter season.
We select the wind direction records from November 1 to November 7, 2007, as the first week data, and the records from January 25 to January 31, 2008, as the last week data. The missing records in two weeks are discarded.

\begin{CodeInput}
R> library("Ball")
R> data("speed.wind", package = "NPCirc")
R> index1 <- which(speed.wind[["Year"]] == 2007 &
+    speed.wind[["Month"]] == 11 & speed.wind[["Day"]] %in% 1:7)
R> index2 <- which(speed.wind[["Year"]] == 2008 &
+    speed.wind[["Month"]] == 1 & speed.wind[["Day"]] %in% 25:31)
R> d1 <- na.omit(speed.wind[["Direction"]][index1])
R> d2 <- na.omit(speed.wind[["Direction"]][index2])
\end{CodeInput}

Each wind direction is one-to-one transformed to a two-dimensional point in the Cartesian coordinates, and then, the difference of any two points is measured by the great-circle distance which is programmed in the \code{nhdist} function in the \pkg{Ball} package.

\begin{CodeInput}
R> theta <- c(d1, d2) / 360
R> dat <- cbind(cos(theta), sin(theta))
R> dx <- nhdist(dat, method = "geo")
\end{CodeInput}

In the final step, we pass the distance matrix and the sample size of two groups to the arguments \code{x} and \code{size}, and set \code{distance = TRUE} to declare that the object passed to the arguments \code{x} is a distance matrix.

\begin{CodeInput}
R> size_vec <- c(length(d1), length(d2))
R> bd.test(x = dx, size = size_vec, distance = TRUE)
\end{CodeInput}
\begin{CodeOutput}
    2-sample Ball Divergence Test

data:  dx
number of observations = 335, group sizes: 168 167
replicates = 99
bd = 0.29114, p-value = 0.01
alternative hypothesis: distributions of samples are distinct
\end{CodeOutput}

\begin{figure}[ht]
\begin{center}
\includegraphics{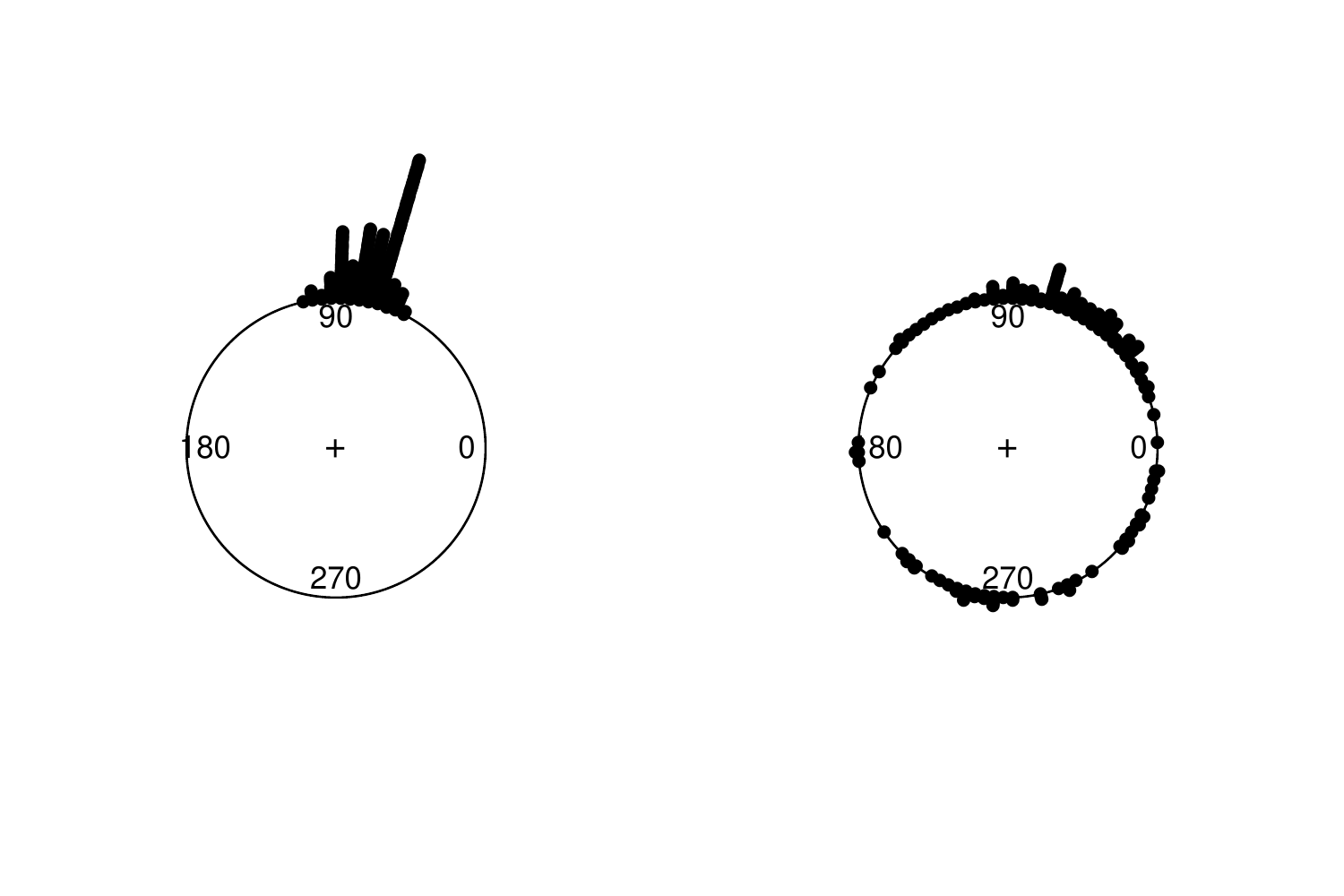}
\end{center}
\vspace{-65.0pt}
\caption{The raw circular data plot of the hourly wind direction dataset. The left panel is corresponding to the first-week wind directions in 2007-08 winter and the right panel is corresponding to the last week's.}\label{wind}
\end{figure}

As can be seen from the output information of \code{bd.test}, \(\stbd\) is 0.2911 and the $p$~value is 0.01. Consequently, at the usual significance level of 0.05, we should reject the null hypothesis. To further confirm our conclusion, we visualize the wind direction of two groups in Figure \ref{wind} with the \proglang{R} package \pkg{circular} \citep{circular2017agostinelli}. Figure \ref{wind} shows that the hourly wind directions in the first week is concentrated around the 90 degrees but the wind directions of the last week are widely dispersed.
\begin{CodeInput}
R> library("circular")
R> par(mfrow = c(1, 2), mar = c(0, 0, 0, 0))
R> plot(circular(c(d1), units = "degrees"), bin = 100, stack = TRUE, 
+    shrink = 1.5)
R> plot(circular(c(d2), units = "degrees"), bin = 100, stack = TRUE,
+    shrink = 1.5)
\end{CodeInput}

\subsubsection{Brain fMRI dataset}\label{independence-test-for-brain-fMRI-data}

We examine a public fMRI dataset from the 1000 Functional Connectomes Project\footnote{\url{https://www.nitrc.org/frs/?group_id=296}}. This project calls on the principal investigators from the member site to donate neuroimaging data such that the broader imaging community have complete access to a large-scale functional imaging dataset. Given the resting-state fMRI and demographics of 86 individuals donated from ICBM, it is
of interest to evaluate whether age is associated with brain connectivity. To properly analyze the dataset, we carry out a preprocessing for the three fMRI of each individual with the \pkg{nilearn} \shortcites{abraham2014nilearn}\citep{abraham2014nilearn} package in the \proglang{Python} environment. The preprocessing for each individual includes four steps: 
(i) segment brain into a set of 116 cortical and subcortical regions for each fMRI with the Automated Anatomical Labeling template \shortcites{tzourio2002automated}\citep{tzourio2002automated}; (ii) average the voxel-specific time series in each of these regions to form mean regional time series for each fMRI; 
(iii) compute the \(116 \times 116\) Pearson correlation coefficient matrix of each fMRI with the 116 mean regional time series, where the Pearson correlation coefficient is a widely-used association measure in the neuroimaging literature \citep{ginestet2017hypothesis,ginestet2011statistical,bullmore2009complex};
(iv) average the three \(116 \times 116\) matrices of each observation in an element-wise manner and save the averaged matrix to disk such that it can be analyzed with \proglang{R}.
In the \proglang{R} environment, the collection of the averaged matrix and demographics are combined into a \code{list} object, then it is saved to a disk as ``niICBM.rda'' file.

To achieve our goal, we compute the pairwise distance matrices of the averaged matrices and the age, then save them in the \proglang{R} objects \code{dx} and \code{dy}, respectively. Then, we pass \code{dx} and \code{dy} to \code{bcov.test} to perform an independence test, meanwhile, let \code{distance = TRUE} to declare what the arguments \code{x} and \code{y} accepted are distance matrices. The detailed \proglang{R} code is demonstrated below.

\begin{CodeInput}
R> library("Ball")
R> library("CovTools")
R> load("niICBM.rda")
R> dx <- as.matrix(CovDist(niICBM[["spd"]]))
R> dy <- as.matrix(dist(niICBM[["covariate"]][["V3"]]))
R> bcov.test(x = dx, y = dy, distance = TRUE, weight = "prob")
\end{CodeInput}

\begin{CodeOutput}
    Ball Covariance test of independence

data:  dx and dy
number of observations = 86
replicates = 99, weight: probability
bcov.probability = 0.017138, p-value = 0.01
alternative hypothesis: random variables are dependent
\end{CodeOutput}

The output message shows that \(\spbcov\) is 0.0171 and the $p$~value is smaller than the usual significance level 0.05, and thus, we conclude that brain connectivity is associated with age. This result is also revealed by recent research that age strongly effects structural brain connectivity \citep{damoiseaux2017effects}. In this example, we use the Euclidean distance to measure the difference between age, and the affine invariant Riemannian metric \citep{pennec2006riemannian} to evaluate the structural difference between the averaged Pearson correlation coefficient matrices. The affine invariant Riemannian metric is implemented in \pkg{CovTools} \citep{covtools2018kyoungjae}

\section{Numerical studies}\label{numerical-study}

In this section, the numerical studies are conducted to assess the performance of the Ball test procedures for complex data, 
including directional data in hyper-sphere spaces, tree-structured data in tree metric spaces, symmetric 
positive definite matrix data in the space of symmetric positive-definite matrices, and functional data in $\mathcal{L}_{\infty}$ spaces. 
Besides, a runtime analysis is provided in Section \ref{runtime-analysis}. For comparison, 
we consider energy distance and HHG for the $K$-sample test problem, while distance covariance, 
distance multivariance, and HHG for the test of mutual independence problem. The permutation technique 
helps us obtain the empirical distributions of these statistics under the null hypothesis, 
and derive their $p$~values. As suggested in \citet{davison1997bootstrap}, at least 99 and at most 999 random permutations
should suffice, and hence, we compute the $p$~value of each test based on 399 random permutations.
All models in Sections \ref{K-sample-test} and \ref{test-of-mutual-independence} are repeated 500 times to estimate Type-I errors and powers. 
In each replication, all methods use the same dataset and the same non-standard distance to 
make a fair comparison. The significance level is fixed at \(0.05\).

\subsection[K-sample test]{$K$-sample test}\label{K-sample-test}

In this section, we investigate the performance of test statistics on revealing the distribution difference with two kinds of complex data, directional data and tree-structured data. They are frequently encountered by scientists interested in wind directions, marine currents \citep{di2014nonparametric}, and cancer evolution \shortcites{abbosh2017phylogenetic}\citep{abbosh2017phylogenetic}. 
To sample directional and tree-structured data, we use the \code{rmovMF} and \code{rtree} functions in the \proglang{R} packages \pkg{movMF} \citep{kurt2014movmf} and \pkg{ape} \citep{paradis2018ape} to draw data from the von Mises-Fisher distribution \(M(\mu, \kappa)\) and random tree distribution \(T(n, \{b_i\}_{i=1}^{n})\), where \(\mu\) and \(\kappa\) are direction and concentration parameters while $n$ and \(\{b_i\}_{i=1}^{n}\) are the numbers of tree nodes and the branch lengths of tree. 
The dissimilarities of two directions and two trees are measured by the great-circle distance and the Kendall Colijn metric \citep{kendall2015tree}, which are programmed in the \code{nhdist} and \code{multiDist} functions in the \proglang{R} packages \pkg{Ball} and \pkg{treespace} \citep{jombart2017treespace}.

We conduct the numerical analyses for directional data in Models 5.1.1, 5.1.3-5.1.5, and 5.1.9 while tree-structured data in other models. Models 5.1.1 and 5.1.2 are designed for Type-I error evaluation, while other models are devoted to evaluating powers. More specifically, Models 5.1.3-5.1.8 focus on the case that any  two groups are different, while Models 5.1.9 and 5.1.10 pay attention to the case that only one group is different to other groups. Without loss of generality, we let \(K=4\) for Models 5.1.1-5.1.8 and \(K=10\) for Models 5.1.9 and 5.1.10. Each group has the same sample size ranging from 10 to 50.
\begin{itemize}
\item{}$\mathrm{Model}$ 5.1.1: von Mises–Fisher distribution. The direction parameters are $\mu^w = \mu^x = \mu^y = \mu^z = (1, 0, 0)$ and the concentration parameters are $\kappa^{w} = \kappa^{x} = \kappa^y = \kappa^z = 30$.
\item{}$\mathrm{Model}$ 5.1.2: Random tree distribution with fifteen nodes. $b^{w}_i, b^{x}_i, b^{y}_i, b^{z}_i, i=1, \ldots, 15$ are independently sampled from the uniform distribution $U(0, 1)$.
\item{}$\mathrm{Model}$ 5.1.3: von Mises–Fisher distribution. The direction parameters are $\mu^w = (0,1, 1, 1, 1)$, $\mu^x = (2, 1, 1, 1, 1), \mu^y = (4, 1, 1, 1, 1), \mu^z = (6, 1, 1, 1, 1)$ and the concentration parameters are $\kappa^{w} = 1, \kappa^{x} = 2, \kappa^{y} = 3, \kappa^z = 4$.
\item{}$\mathrm{Model}$ 5.1.4: Mixture von Mises–Fisher distribution. 
The direction parameters are $\mu^{w_1} = \mu^{x_1} = (1, 0), \mu^{w_2} = \mu^{x_2} = (-1, 0), \mu^{y_1} = \mu^{z_1} = (0, 1), \mu^{y_2} = \mu^{z_2} = (0, -1)$ 
and the concentration parameters are $\kappa^{w_1}=\kappa^{w_2}=\kappa^{y_1}=\kappa^{y_2}=30, \kappa^{x_1}=\kappa^{x_2}=\kappa^{z_1}=\kappa^{z_2}=35$. 
The four mixture proportions of two von Mises–Fisher distributions are all 0.5.
\item{}$\mathrm{Model}$ 5.1.5: Mixture von Mises–Fisher distribution. The direction parameters are $\mu^{w_1} = (1, 0, 0, 0), \mu^{w_2} = (-1, 0, 0, 0), \mu^{x_1} = (0, 1, 0, 0), \mu^{x_2} = (0, -1, 0, 0), \mu^{y_1} = (0, 0, 1, 0)$, $\mu^{y_2} = (0, 0, -1, 0), \mu^{z_1} = (0, 0, 0, 1), \mu^{z_2} = (0, 0, 0, -1)$ and the concentration parameters are all 30. The four mixture proportions of two von Mises–Fisher distributions are all 0.5.
\item{}$\mathrm{Model}$ 5.1.6: Random tree distribution with fifteen nodes. $b^{w}_i, b^{x}_i, b^{y}_i, b^{z}_i, i = 1, \ldots, 15$ are independently sampled from four different uniform distributions: $b^{w}_i \sim U(0, 0.25), b^{x}_i \sim U(0, 0.5), b^{y}_i \sim U(0, 0.75), b^{z}_i \sim U(0, 1), i = 1, \ldots, 15$.
\item{}$\mathrm{Model}$ 5.1.7: Random tree distribution with fifteen nodes. $b^{w}_i, b^{x}_i, b^{y}_i, b^{z}_i, i = 1, \ldots, 15$ are independently sampled from four different uniform distributions: $b^{w}_i \sim U(0, 12), b^{x}_i \sim U(2, 10), b^{y}_i \sim U(3, 8), b^{z}_i \sim U(4, 6), i = 1, \ldots, 15$.
\item{}$\mathrm{Model}$ 5.1.8: Random tree distribution with fifteen nodes. $b^{w}_i, b^{x}_i, b^{y}_i, b^{z}_i, i = 1, \ldots, 15$ are independently sampled from four different $F$ distributions: $b^w_i \sim F(2, 2), b^x_i \sim F(2, 3)$, $b^y_i \sim F(2, 4)$, $b^z_i \sim F(2, 5), i = 1, \ldots, 15$.
\end{itemize}
Since only one group is different to other groups, it is sufficient to specify the following Models by describing the distributions of two groups.
\begin{itemize}
\item{}$\mathrm{Model}$ 5.1.9: von Mises–Fisher distribution. The direction parameters are $\mu^x = (0, 1, 1,1,1)$, $\mu^y = (2, 1, 1, 1, 1)$ and the concentration parameters are $\kappa^{x} = \kappa^{y} = 3$.
\item{}$\mathrm{Model}$ 5.1.10: Random tree distribution with fifteen nodes. $b^{x}_i, b^{y}_i, i = 1, \ldots, 15$ are independently sampled from two different uniform distribution: $b^{x}_i \sim U(0, 0.5), b^{y}_i \sim U(0, 1), i = 1, \ldots, 15$.
\end{itemize}

\begin{figure}[ht]
\begin{center}
\includegraphics[scale=1.0]{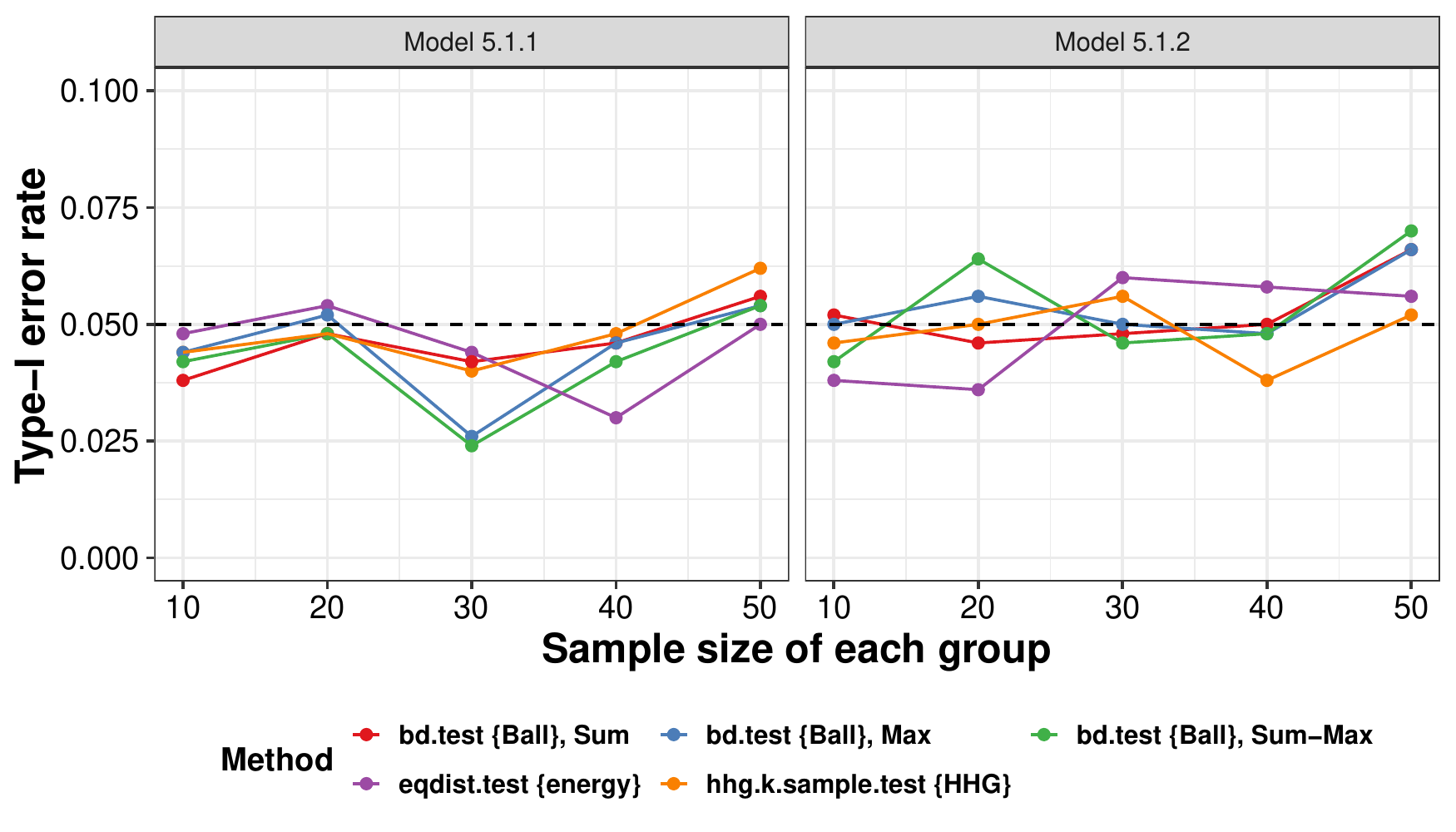}
\end{center}
\vspace{-20.0pt}
\caption{Type-I error rates of the five tests for the $K$-sample test problem. The black dashed line is the nominal significance level.}\label{simulate_tab1}
\end{figure}

\begin{figure}[ht]
\begin{center}
\includegraphics[scale=1.0]{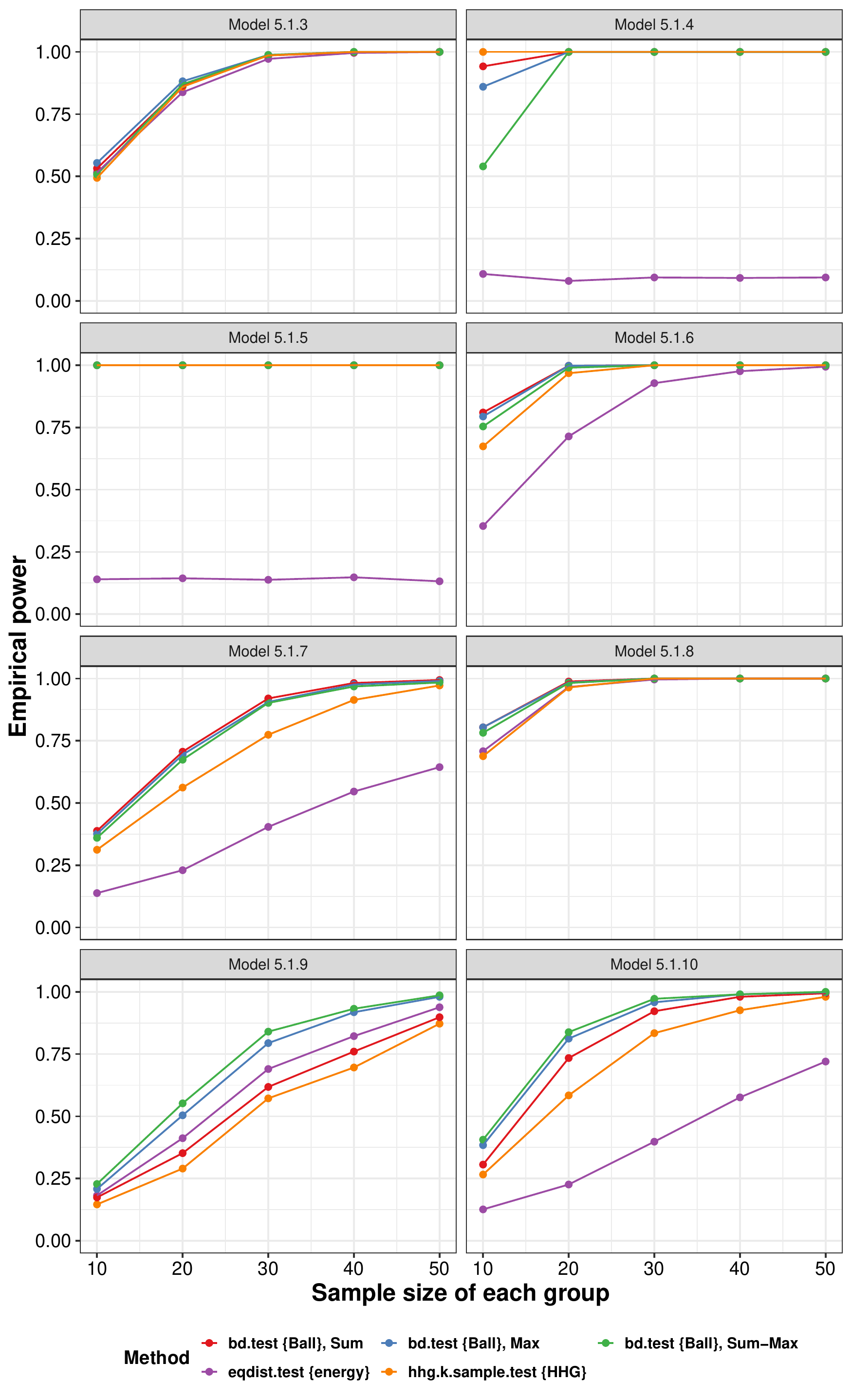}
\end{center}
\vspace{-20.0pt}
\caption{Power estimates of the five tests for the $K$-sample test problem.}\label{simulate_tab2}
\end{figure}

The Type-I error rates and power estimates are demonstrated in Figures \ref{simulate_tab1} and \ref{simulate_tab2}. 
From Figure~\ref{simulate_tab1}, all test methods can control the Type-I error rates well around the significance level. 
Figure~\ref{simulate_tab2} shows that \(\sksbd, \skmsbd,\) or \(\skmbd\) outperforms energy distance and HHG in most cases. 
More specifically, \(\sksbd\) is generally superior to other methods when any two groups are different, 
and \(\skmsbd\) has an advantage when the relatively rare group distinctions increase the difficult of the $K$-sample test problem.
As for \(\skmbd\), it is more stable compared with \(\sksbd\) and \(\skmsbd\). 
\(\skmbd\) is better than \(\skmsbd\) when any two groups are different, and better than \(\sksbd\) when the group distinctions are relatively rare.

In Models 5.1.4 and 5.1.5, the empirical powers of energy distance are low and not increasing, yet the $K$-sample BD statistics and HHG are well-performed. This is not surprising because the great-circle distance is not of strong negative type.
We provide an example to illustrate this result as follows. 
Without loss of generality, we simplify the distributions in Model 5.1.4 to the equal-probability Bernoulli distributions on a circle, where $W, X$ take $(0, 1)$ and $(0, -1)$ while $Y, Z$ take $(1, 0)$ and $(-1, 0)$. Then, the $K$-sample test problem could be considered as the two-sample test problem between groups $X$ and $Y$. For the two groups $X$ and $Y$, the means of the great-circle distance within the group are both $\pi/2$, so is the mean of the great-circle distance between the two groups. The energy distance between $X$ and $Y$ is zero according to its definition \citep{szekely2013energy}
\begin{equation}\label{ed_definition}
\mathcal{E}(X, Y) = 2E(d(X, Y)) - E(d(X, X^{\prime}))- E(d(Y, Y^{\prime})),
\end{equation}
where $X^{\prime}$ and $Y^{\prime}$ are \emph{i.i.d.} copy of $X$ and $Y$, respectively. 
Thus, energy distance fails to detect the distribution difference. On the contrary, $\stbd$ is larger than zero since all observations in $B((0, 1), (0, 1)) \cup B((0, -1), (0, -1))$ come from the group $X$ and all observations in $B((1, 0), (1, 0))\cup B((-1, 0), (-1, 0))$ come from the group $Y$. The result of Model 5.1.5 can be interpreted similarly.

\subsection{Test of mutual independence}\label{test-of-mutual-independence}

In this section, we evaluate the performance of test methods on detecting the relationship among complex random objects.
The complex random objects attracting our attention are symmetric positive definite matrix and functional curve, commonly encountered in contemporary statistical research, for instance, \citet{dryden2009noneuclidean} and \citet{wang2016functional}. We generate the two types of random objects with the \code{genPositiveDefMat} function in the \proglang{R} package \pkg{clusterGeneration} \citep{qiu2015clusterGeneration} and a series of functions in the \proglang{R} package \pkg{fda} \citep{ramsay2018fda}. The \code{genPositiveDefMat} function can generate a random \(d \times d\) symmetric positive definite matrix \(SPD_d(\lambda, \rho) (\lambda>0, \rho>1)\) whose eigenvalues range from \(\lambda\) to \(\rho \lambda\). The \proglang{R} functions in \pkg{fda} help construct the functional curve $f(t; \coef)=\coef^{\top}(1, \sin{t}, \cos{t})$ and acquire 17 observed points which are equally spaced at interval $[0, 8\pi]$, where $\coef$ is a coefficient vector. The typical dissimilarity measurements of two symmetric positive definite matrices and two functional curves are the affine invariant Riemannian metric and the $L_{\infty}$ norm, which are implemented in the \proglang{R} packages \pkg{CovTools} and \pkg{fda.usc} \citep{manuel2012fdausc}, respectively.

We design Models 5.2.1-5.2.2 for the examination of Type-I errors and
Models 5.2.3-5.2.10 for the assessment of powers. Let the sample size increase from 40 to 120.
\begin{itemize}
\item{}$\mathrm{Model}$ 5.2.1: $X, Y$ are independently sampled from the $SPD_{10}(1, 10).$
\item{}$\mathrm{Model}$ 5.2.2: $\coef_{1}, \coef_{2}$ are independently sampled from the multivariate uniform distribution on the cube $[0, 1]^3,$
\begin{align*}
X(t) = f(t; \coef_{1}), Y(t) = f(t; \coef_{2}).
\end{align*}
\item{}$\mathrm{Model}$ 5.2.3: $Z$ comes from the uniform distribution $U(0, \pi/2)$, and $\epsilon_{1}, \epsilon_{2}, \epsilon_{3}$ are independently sampled from the Chi-square distribution with the degree of freedom 1,
\begin{align*}
X \sim SPD_{10}\left(Z, 1 + \epsilon_1 \right), Y \sim SPD_{10}\left(Z + \epsilon_2, 1 + \epsilon_3 \right).
\end{align*}
\item{}$\mathrm{Model}$ 5.2.4: The distributions of $Z, \epsilon_{1}, \epsilon_{2}, \epsilon_{3}$ are the same as in Model 5.2.3.
\begin{align*}
X \sim SPD_{10}\left(Z, 1 + \epsilon_1 \right), Y \sim SPD_{10}\left( 10 \left|\cos{(3Z)}\right| + \epsilon_2, 1 + \epsilon_3 \right).
\end{align*}
\item{}$\mathrm{Model}$ 5.2.5: $\coef_1, \coef_2, \coef_3$ are independent standard normal random vectors, and $\epsilon_{1}(t)$ and $\epsilon_{2}(t)$ are independent gaussian processes,
\begin{align*}
&Z_1(t) = f(t; \coef_1), Z_2(t) = f(t; \coef_2), Z_3(t) = f(t; \coef_3), \\
X(t) = 4Z_1(t&)Z_2(t)Z_3(t) + \epsilon_{1}(t), Y(t) = [Z_{1}(t) - Z_{2}(t)]^2 + 2Z_{3}(t) + \epsilon_{2}(t).
\end{align*}
\item{}$\mathrm{Model}$ 5.2.6: $X$ comes from the binomial distribution $B(1, 0.5)$, and $\epsilon(t)$ is a gaussian process,
\begin{align*}
P(\coef = (0, 0, 1)|X = 0) &= P(\coef = (0, 0, -1)|X = 0) = 0.5,\\
P(\coef = (0, 1, 0)|X = 1) &= P(\coef = (0, -1, 0)|X = 1) = 0.5,\\
Y(t) = 1&0f(t; \coef) + \epsilon(t).
\end{align*}
\end{itemize}
The following four models are constructed for evaluating the power of test methods in the test of mutual independence problem. To the best of our knowledge, only \pkg{multivariance} and \pkg{Ball} allow \proglang{R} users to perform the mutual independence test on datasets in metric spaces.
And hence, we only compare \pkg{Ball} and \pkg{multivariance} below.
\begin{itemize}
\item{}$\mathrm{Model}$ 5.2.7: The distributions of $Z, \epsilon_{1}, \epsilon_{2}, \epsilon_{3}$ are the same as Model 5.2.3, and $\epsilon_{4}, \epsilon_{5}$ are independently drawn from Pareto distribution with location parameter 0.8 and shape parameter 1.
\begin{align*}
&X \sim SPD_{10}\left(Z, 1 + \epsilon_1 \right), \\
Y &\sim SPD_{10}\left(Z + \epsilon_4, 1 + \epsilon_2 \right),\\
Z &\sim SPD_{10}\left(Z + \epsilon_5, 1 + \epsilon_3 \right).
\end{align*}
\item{}$\mathrm{Model}$ 5.2.8: $W_1, W_2$ are independently drawn from the $t$ distribution with the degree of freedom 1,
\begin{align*}
&\quad \; X \sim SPD_{10}(1, |W_1| + 1), \\
&Y \sim SPD_{10}(1, 1 + (W_1 - W_2)^2), \\
Z \sim& SPD_{10}(1, \exp{(8\sin(W_1 - W_2))} + 1).
\end{align*}
\item{}$\mathrm{Model}$ 5.2.9: $Z_1, Z_2$ are independently sampled from the binomial distribution $B(1, 0.5)$, $Z_3 = I(Z_1 = Z_2)$. Let $\coef_0 = (0, 1, 0), \coef_1 = (0, 0, 1)$,
\begin{align*}
X(t) = I(Z_1=0)f(t; \coef_0) + I(Z_1=1)f(t; \coef_1), \\
Y(t) = I(Z_2=0)f(t; \coef_0) + I(Z_2=1)f(t; \coef_1), \\
Z(t) = I(Z_3=0)f(t; \coef_0) + I(Z_3=1)f(t; \coef_1).
\end{align*}
\item{}$\mathrm{Model}$ 5.2.10: $X$ is sampled from the binomial distribution $B(1, 0.5)$, and $\epsilon_1(t)$ and $\epsilon_2(t)$ are independent gaussian processes,
\begin{align*}
P(\coef = (0, 0, 1)|X = 0) &= P(\coef = (0, 0, -1)|X = 0) = 0.5, \\
P(\coef = (0, 1, 0)|X = 1) &= P(\coef = (0, -1, 0)|X = 1) = 0.5, \\
Y(t) = 10f(t; &\coef) + \epsilon_{1}(t), \; Z(t) = \epsilon_{2}(t).
\end{align*}
\end{itemize}

\begin{figure}[ht]
\begin{center}
\includegraphics[scale=1.0]{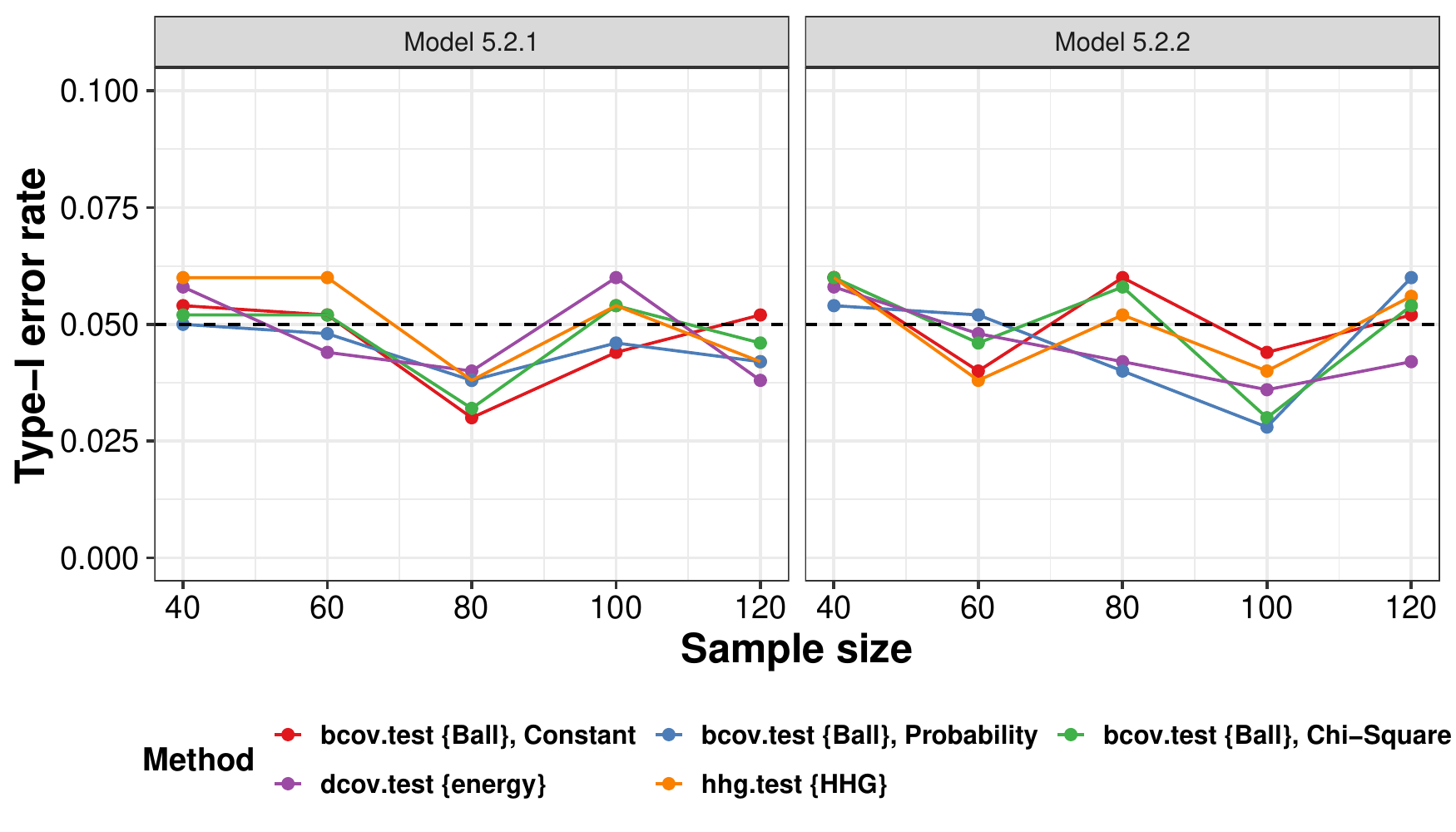}
\end{center}
\vspace{-20.0pt}
\caption{Type-I error rates of the five tests for the test of independence problem. The black dashed line is the nominal significance level.}\label{simulate_tab3}
\end{figure}

\begin{figure}[ht]
\begin{center}
\includegraphics[scale=1.0]{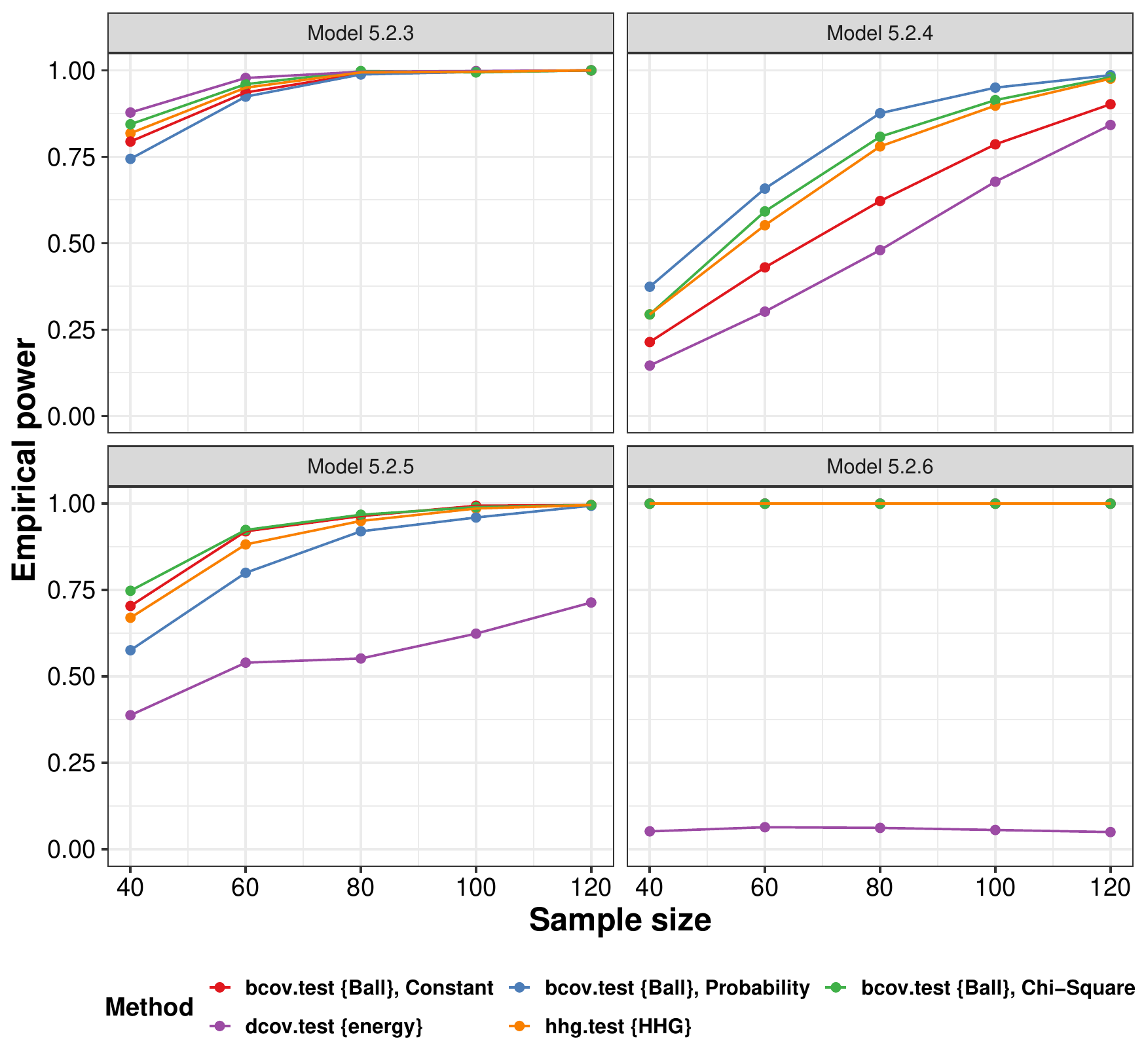}
\end{center}
\vspace{-20.0pt}
\caption{Power estimates of the five tests for the test of independence problem.}\label{simulate_tab4}
\end{figure}

\begin{figure}[ht]
\begin{center}
\includegraphics[scale=1.0]{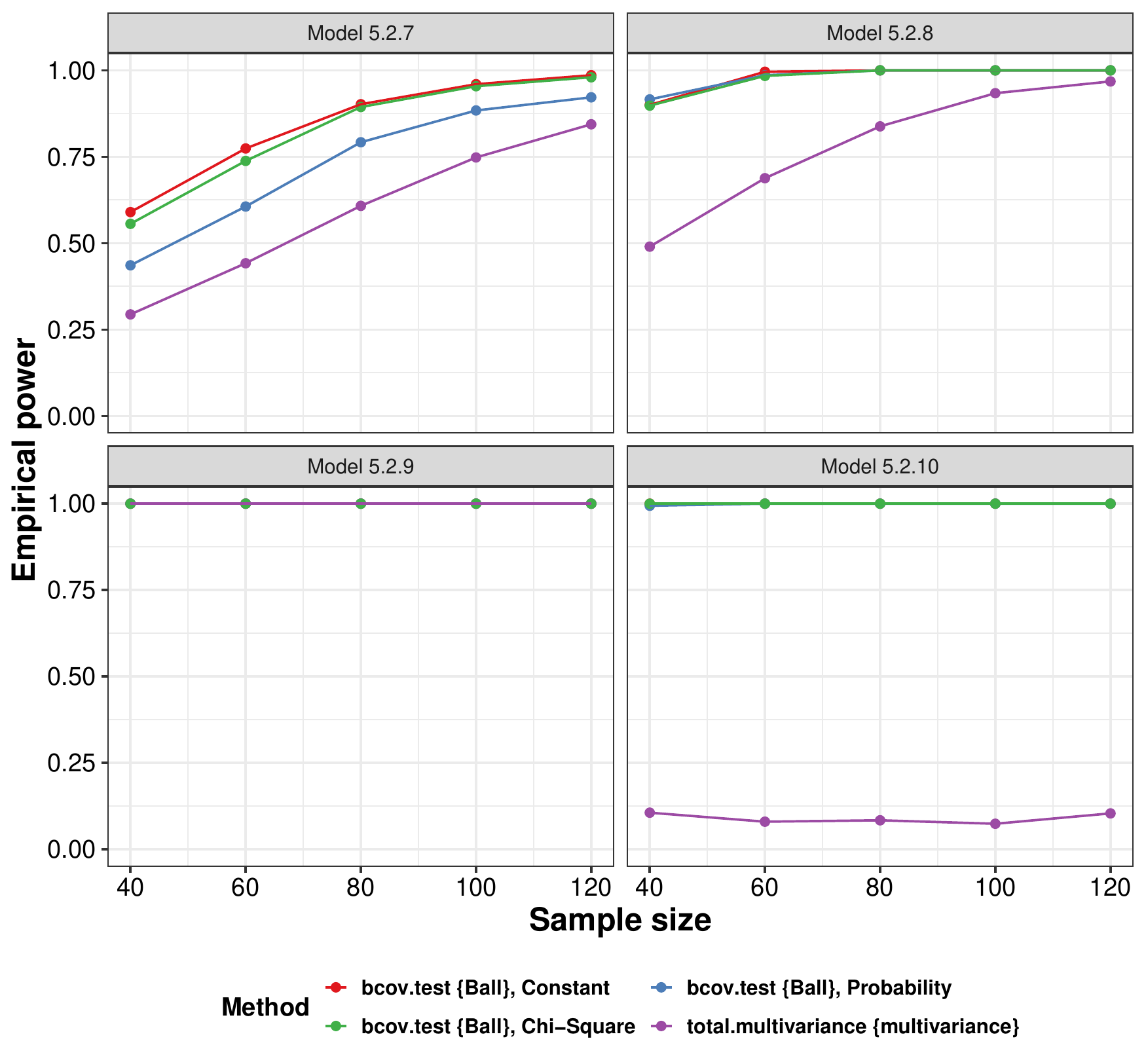}
\end{center}
\vspace{-20.0pt}
\caption{Power estimates of the four tests for the test of mutual independence problem.}\label{simulate_tab5}
\end{figure}

The Type-I error rates and empirical power are displayed in Figures \ref{simulate_tab3}, \ref{simulate_tab4}, and \ref{simulate_tab5}. 
As shown in Figure \ref{simulate_tab3}, the Type-I error rates of all methods are reasonably controlled around the significance level. 
From Figure \ref{simulate_tab4}, both the BCOV statistics and HHG are competitive and generally exceed distance covariance. From Figure \ref{simulate_tab5}, the three BCOV statistics successfully detect the complicated mutual dependence among multiple random objects, and their empirical powers increase as the sample size augments. It is worth noting that Model 5.2.9 is an example of pairwise independence with mutual dependence. The success in revealing the mutual dependence of Model 5.2.9 certifies the power of the BCOV statistics. 

To shed a light on the performance difference of the three BCOV statistics, we compare their empirical powers in Models 5.2.3, 5.2.4, and 5.2.7. In Model 5.2.3, the lower bound of the eigenvalues of $X$ is linearly associated with that of $Y$, and similarly in Model 5.2.7, except that the noise in Model 5.2.7 has an infinite first moment. $\skchibcov$ has a best performance in Model 5.2.3 on account of the nonlinearity of symmetric positive definite matrices spaces which slightly improves the nonlinearity between $X$ and $Y$. In Model 5.2.7, $\skcbcov$ is superior to other methods owing to the approximately linear relationship among random objects and its high robustness. 
As for Model 5.2.4, the lower bounds of the eigenvalues of $X$ and $Y$ have a strongly nonlinear relationship. At this point, $\skpbcov$ turns to be the first place.

It is also worthwhile to take a good look at Models 5.2.6 and 5.2.10. In the two models, the empirical power of distance covariance and distance multivariance stay at a low level as the sample size increases, because the $L_{\infty}$ norm is not of strong negative type. 
The following is an explanation of why distance covariance has an unsatisfactory performance in Model~5.2.6.
Without loss of generality, we re-define $Y(t) = 10f(t; \coef)$, then denote $Y(t)|X = 0$ and $Y(t)|X = 1$ as $Y_1(t)$ and $Y_2(t)$. It is easy to verify that the distance covariance of $(X, Y(t))$ is the constant-multiple energy distance between groups $Y_1(t)$ and $Y_2(t)$. For the two groups $Y_1(t)$ and $Y_2(t)$, the means of the $L_{\infty}$ norm within the group are both $10$, so is the mean of the $L_{\infty}$ norm between the two groups. According to Equation~\ref{ed_definition}, the energy distance between $Y_1(t)$ and $Y_2(t)$ is 0, and thus, the distance covariance of $(X, Y(t))$ is 0, leading to the failure of detecting association. 
The performance of distance multivariate in Model 5.2.10 could be explained similarly.


\subsection{Runtime analysis}\label{runtime-analysis}
\begin{figure}[ht]
\hspace{-0.1\textwidth}
\includegraphics[width=1.15\textwidth]{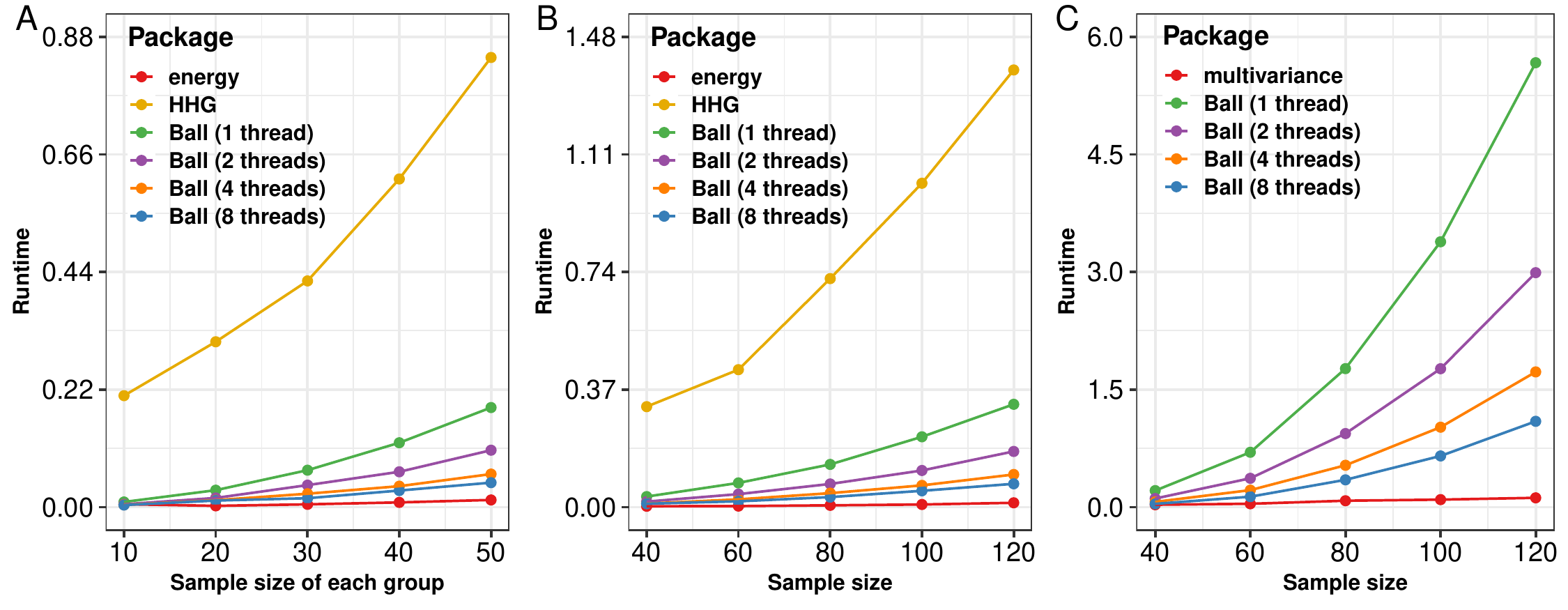}
\vspace{-20.0pt}
\caption{Runtime analysis for \pkg{energy}, \pkg{multivariance}, \pkg{HHG}, and \pkg{Ball} in the $K$-sample test and test of (mutual) independence problems in metric spaces. A) the $K$-sample test problem ($K = 4$), B) the test of independence problem, C) the test of mutual independence problem. The runtimes ($y$-axis) are measured in seconds.}\label{runtime_tab6}
\end{figure}

We adopt Models 5.1.1, 5.2.1 and 5.2.7 in Sections \ref{K-sample-test} and \ref{test-of-mutual-independence} to assess the runtime performance of \pkg{energy} (1.7.5), \pkg{multivariance} (2.1.0), \pkg{HHG} (1.3.2), and \pkg{Ball} (1.3.8) using the \pkg{microbenchmark} package \citep{olaf2018microbenchmark}. Here, all experiments are conducted with 20 replications,
and the averaged runtimes are visualized in Figure \ref{runtime_tab6}. The benchmark is a 64-bit Windows platform with Intel Core i7 @ 3.60 GHz.

From Figure \ref{runtime_tab6}, \pkg{energy} is the fastest package in the $K$-sample test and the test of independence problems, and \pkg{multivariance} is the fastest package in the test of mutual independence problem. As the second fastest package, \pkg{Ball} is around four times faster than \pkg{HHG} in the $K$-sample test and test of independence problems when both of them use one thread, even though $\schibcov(K = 2)$ and HHG are asymptotically
equivalent. Furthermore, we can cut the runtimes of \pkg{Ball} down around one third via doubling threads.

In summary, if runtimes are more concerned, \pkg{energy} or \pkg{multivariance} may be a desirable choice. Otherwise, \pkg{Ball} is a preferable choice due to its powerful performance in various complex data with fewer runtime increase, especially for the $K$-sample test and the test of independence problems.

\section{Conclusion}\label{conclusion}

We design a user-friendly \proglang{R} package \pkg{Ball} to help data scientists detect the distribution distinction and object association for complex data in metric spaces. Equipped with the novel algorithms, efficient \proglang{C} implementation, advanced multi-threaded technique, and elegant theoretical properties of the Ball test statistics, the Ball test procedures programmed in the \pkg{Ball} package can efficiently analyze complex data in metric spaces.

Future versions of the \pkg{Ball} package will endeavor to speed up the Ball Correlation based generic feature screening procedure \citep{pan2018generic}. Furthermore, we intend to develop \proglang{Python} and \proglang{Julia} packages to help data scientists conduct the Ball test procedures and Ball screening procedure with their most familiar program languages.

\section*{Acknowledgment}
We would like to thank referees for their valuable comments and suggestions which have substantially improved this article. 

Dr.~Pan's research is partially supported by the National Natural Science Foundation of China (11701590), Natural Science
Foundation of Guangdong Province of China (2017A030310053) and Young teacher program/Fundamental Research Funds for the Central
Universities (17lgpy14). Dr.~Zheng's research is partially supported by National Science Foundation, DMS-1830864. 
Dr.~Wang's research is partially supported by NSFC (11771462), 
International Science \& Technology cooperation program of Guangdong, China (2016B050502007), 
The National Key Research and Development Program of China (2018YFC1315400), 
and The Key Research and Development Program of Guangdong, China (2019B020228001).

\bibliography{reference}

\section*{Appendix}\label{appendix}
Merge sort is a classical divide-and-conquer algorithm for sorting. 
It recursively splits the value array in half until all subarrays only have one element, 
then merges those subarrays to a sorted array. To adapt to the ``count of the smaller number after 
self'' problem, merge sort uses an auxiliary equal-size number array to record the numbers of the smaller 
element after self. Initialized all elements with 0, the number array is split and merged 
with the value array. In the merging stage, if an element of the left side value array is to be merged, then, 
the merged elements of the right side value array must be no larger than the element to be merged. And hence, 
the corresponding element in the left side number array should add the number of the merged elements of the right side value array. 
Implemented with \proglang{C} in the \pkg{Ball} package, the solution of ``count of the numbers after self'' problem is given below.
\begin{Code}
C> void count_smaller_number_after_self_solution(double *value, int *number, 
+  					    const int num) {
+      int index[num];
+      for (int i = 0; i < num; ++i) {
+          index[i] = i;
+      }
+      merge_sort(value, index, number, 0, num - 1);
+  }
C> void merge_sort(double *value, int *index, int *number, int start, int end) {
+      if (end - start < 1) return;
+      int mid = (start + end) >> 1;
+      merge_sort(value, index, number, start, mid);
+      merge_sort(value, index, number, mid + 1, end);
+      merge(value, index, number, start, mid, end);
+  }
C> void merge(double *value, int *index, int *number, int start, int mid, int end) {
+      const int left_size = mid - start + 1, right_size = end - mid;
+      double left[left_size], right[right_size];
+      int left_index[left_size], right_index[right_size];
+      int left_merged = 0, right_merged = 0, total_merged = 0;
+      for (int i = start; i <= mid; ++i) {
+          left[i - start] = value[i];
+          left_index[i - start] = index[i];
+      }
+      for (int i = mid + 1; i <= end; ++i) {
+          right[i - mid - 1] = value[i];
+          right_index[i - mid - 1] = index[i];
+      }
+      while (left_merged < left_size && right_merged < right_size) {
+          if (left[left_merged] < right[right_merged]) {
+              number[left_index[left_merged]] += right_merged;
+              value[start + total_merged] = left[left_merged];
+              index[start + total_merged] = left_index[left_merged];
+              ++left_merged;
+              ++total_merged;
+          } else {
+              value[start + total_merged] = right[right_merged];
+              index[start + total_merged] = right_index[right_merged];
+              ++right_merged;
+              ++total_merged;
+          }
+      }
+      while (left_merged < left_size) {
+          number[left_index[left_merged]] += right_merged;
+          value[start + total_merged] = left[left_merged];
+          index[start + total_merged] = left_index[left_merged];
+          ++left_merged;
+          ++total_merged;
+      }
+      while (right_merged < right_size) {
+          value[start + total_merged] = right[right_merged];
+          index[start + total_merged] = right_index[right_merged];
+          ++right_merged;
+          ++total_merged;
+      }
+  }
\end{Code}
\end{document}